\documentclass[aps,prb,twocolumn,showpacs,amsmath,amssymb,superscriptaddress,floatfix]{revtex4-2}

\usepackage[english]{babel}
\usepackage{blindtext}
\usepackage{latexsym}
\usepackage{amssymb}
\usepackage{physics}
\usepackage{amsmath}
\usepackage{bm}
\usepackage{mathtools}
\usepackage{amsfonts}
\usepackage{relsize}
\usepackage{xcolor}
\usepackage{verbatim}
\usepackage{lipsum}  
\usepackage{bbold}
\usepackage{slashed}
\usepackage{appendix} 
\usepackage{graphicx}
\usepackage[normalem]{ulem}
\usepackage[colorlinks = true,
            linkcolor = blue,
            urlcolor  = blue,
            citecolor = blue,
            anchorcolor = blue]{hyperref}
\usepackage{graphicx} 
\usepackage{dcolumn} 
\newcommand{\approptoinn}[2]{\mathrel{\vcenter{
  \offinterlineskip\halign{\hfil$##$\cr
    #1\propto\cr\noalign{\kern2pt}#1\sim\cr\noalign{\kern-2pt}}}}}

\newcommand*{\colorboxed}{}
\def\colorboxed#1#{%
  \colorboxedAux{#1}%
}
\newcommand*{\colorboxedAux}[3]{%
  \begingroup
    \colorlet{cb@saved}{.}%
    \color#1{#2}%
    \boxed{%
      \color{cb@saved}%
      #3%
    }%
  \endgroup
}

\makeatletter
\renewcommand*\env@matrix[1][\arraystretch]{%
  \edef\arraystretch{#1}%
  \hskip -\arraycolsep
  \let\@ifnextchar\new@ifnextchar
  \array{*\c@MaxMatrixCols c}}
\makeatother

\newcommand{\bs}[1]{\mathbf{#1}}

\newcommand{\bq}{\bs{q}}
\newcommand{\bQ}{\bs{Q}}
\newcommand{\bzero}{\bs{0}}
\newcommand{\bk}{\bs{k}}

\newcommand{\eps}{\epsilon}

\definecolor{ao(english)}{rgb}{0.0, 0.5, 0.0}
\definecolor{amaranth}{rgb}{0.9, 0.17, 0.31}
\definecolor{green(html/cssgreen)}{rgb}{0.0, 0.5, 0.0}

\newcommand\greensout{\bgroup\markoverwith{\textcolor{green(html/cssgreen)}{\rule[0.5ex]{2pt}{1.0pt}}}\ULon}

\begin{document}
\title{Coexisting magnetic, charge, and superconducting orders \\
in the two-dimensional Hubbard model}

%
\author{Robin Scholle}
\affiliation{Max Planck Institute for Solid State Research, D-70569 Stuttgart, Germany}
\author{Pietro M.~Bonetti}
\affiliation{Department of Physics, Harvard University, Cambridge, Massachusetts 02138, USA}
\affiliation{Max Planck Institute for Solid State Research, D-70569 Stuttgart, Germany}
\author{Walter Metzner}
\affiliation{Max Planck Institute for Solid State Research, D-70569 Stuttgart, Germany}
\author{Demetrio Vilardi}
\affiliation{Max Planck Institute for Solid State Research, D-70569 Stuttgart, Germany}
\date{\today}

\begin{abstract}
We perform a renormalized mean-field study of the two-dimensional repulsive Hubbard model, focusing on the intricate interplay and possible coexistence of magnetic, charge, and superconducting orders. We improve on conventional mean-field theory by utilizing a renormalization group framework that captures high-energy fluctuations. This method generates effective magnetic and $d$-wave pairing interactions, and allows for an unbiased exploration of coexisting phases at weak and moderate interaction strengths. Unrestricted mean-field calculations of the effective Hamiltonian on large finite lattices are combined with analyses in the thermodynamic limit, revealing a rich phase diagram with extensive regions of coexisting orders. We find that $d$-wave superconductivity coexists with N\'eel order on the electron-doped side. On the hole-doped side, superconductivity is found to coexist with spiral or stripe magnetic orders. Within the stripe ordered region, the superconducting order parameter is spatially modulated, with a period that follows the charge modulation of the stripes. Below van Hove filling, pairing provides the primary energy gain, while the stripe order yields only a small, and hence fragile, additional energy lowering.
\end{abstract}
\maketitle


\section{Introduction}

The Hubbard model is a prototypical lattice fermion model for correlated electrons. Its two-dimensional version on the square lattice captures much of the essential physics of electrons in the copper-oxide planes of cuprate high temperature superconductors \cite{Scalapino2012}.
The complexity of the Hubbard model arises from the competition between the kinetic energy and the on-site repulsion, which gives rise to a variety of competing phases, including magnetism, charge order, and unconventional superconductivity.
In spite of significant efforts and progress, only fragments of the phase diagram of the two-dimensional Hubbard model, as a function of hopping amplitudes, coupling strengths, electron density, and temperature, have been firmly established \cite{Arovas2022, Qin2022}.

Magnetic order in the two-dimensional Hubbard model has been studied for decades. While a N\'eel antiferromagnetic state is well-established at half-filling, the scenario at finite doping is quite rich -- and still under debate.
Most calculations yield either planar spin spirals \cite{Shraiman1989, Shraiman1992, Chubukov1992, Chubukov1995, Dombre1990, Fresard1991, Kampf1996, Kotov2004, Igoshev2010, Igoshev2015, Yamase2016, Eberlein2016, Mitscherling2018, Vilardi2020, Bonetti2020a}, or spin-charge stripes with collinear spin order and concomitant charge density waves \cite{Schulz1989, Zaanen1989, Machida1989, Poilblanc1989, Schulz1990, Kato1990, Seibold1998, Fleck2000, Fleck2001, Kivelson2003, Raczkowski2010, Timirgazin2012, Peters2014, Huang2018, Matsuyama2022, Zheng2017, Qin2020}, as energetically most favorable.
An unbiased magnetic phase diagram involving N\'eel, spiral, and stripe order has been constructed only recently, but only at weak coupling within unrestricted Hartree-Fock theory \cite{Scholle2023, Scholle2024}.

The possibility of $d$-wave pairing adds another level of complexity to the problem. Magnetic fluctuations generate $d$-wave pairing, while magnetic order competes with $d$-wave superconductivity. However, magnetism and superconductivity may also coexist. Coexistence of $d$-wave superconductivity with magnetic N\'eel order in the Hubbard model has been found, for example, in Refs.~\cite{Lichtenstein2000, Aichhorn2006, Capone2006, Reiss2007, Kancharla2008, Wang2014}, and with spiral order in Refs.~\cite{Yamase2016, Bonetti2020a}.
Recently, coexistence between stripe order and superconductivity has been found in the ground state at various choices of the electron density~\cite{Xu2024}.

Magnetic and charge order can be studied most easily within the Hartree-Fock mean-field approximation. This method can be applied directly in the thermodynamic limit by restricting the order to ferromagnetic, N\'eel \cite{Lin1987, Hofstetter1998, Langmann2007}, or spiral states \cite{Kampf1996,Igoshev2010}. Alternatively, generic magnetic and charge orders can be captured by using an unrestricted Hartree-Fock theory on finite lattices \cite{Zaanen1989, Seibold1998, Fleck2001, Scholle2023}. Although Hartree-Fock theory generally overestimates the stability of ordered states, it provides a valuable guide for more advanced studies.

However, superconductivity in the (repulsive) Hubbard model is fluctuation driven, and therefore cannot be captured by plain mean-field theory.
To overcome this limitation, one may use a renormalized mean-field theory \cite{Wang2014}, where high-energy degrees of freedom are integrated out via the functional renormalization group \cite{Metzner2012Review}. This procedure leads to an effective non-local pairing interaction capable of giving rise to $d$-wave superconductivity. This method has already been used with a restriction of magnetic order to N\'eel or spiral, yielding coexistence of magnetism and superconductivity in a broad region of the phase diagram \cite{Wang2014, Yamase2016, Vilardi2020}.

In this paper, we present an extensive renormalized mean-field study of the competition and coexistence of magnetism, charge order, and superconductivity in the two-dimensional Hubbard model at moderate interaction strength. The spin and charge order pattern is thereby not restricted a priori, but obtained without any bias by minimizing the free energy.
Renormalized unrestricted mean-field calculations on large finite lattices are complemented by analyses in the thermodynamic limit. Our results show that N\'eel, spiral, and stripe orders are the dominant magnetic phases at intermediate interaction strengths. We find that these magnetic phases can coexist with $d$-wave superconductivity at low temperatures. Stripe order induces a modulation of the superconducting amplitude that follows the charge modulation of the stripes.

The renormalized mean-field theory yields phases with spontaneously broken continuous symmetries and corresponding long-range order even at finite temperatures, violating thus the Mermin-Wagner theorem \cite{Mermin1966}. This is because order parameter fluctuations are missing in this approach.
However, the results from such a mean-field approach are still very useful in the context of electron fractionalization, where electrons are described, for example, in terms of a fermionic chargon field and a spinon field that captures magnetic fluctuations \cite{Scheurer2018, Sachdev2019review, Bonetti2022gauge}. In this framework, a renormalized mean-field calculation can serve as a description of the chargons, and the spinon fluctuations restore the broken spin symmetry.
Similarly, the superconducting transition obtained in the mean-field theory actually describes only the formation of Cooper pairs, while the true (Kosterlitz-Thouless) superconducting transition occurs at lower temperatures, and can be computed from the superfluid stiffness obtained from the mean-field theory \cite{Vilardi2020}.

The paper is structured as follows: In Sec.~\ref{Sec: ModelMethod}, we introduce the renormalized mean-field procedure, and a classification scheme of the magnetic states we find. We then present our numerical results in Sec.~\ref{Sec: NumericalResults}. Combining results obtained on large finite lattices with results obtained directly in the thermodynamic limit, we present a comprehensive phase diagram of the Hubbard model in filling and temperature, at a fixed moderate interaction strength. Finally, we conclude with a summary and outlook in Sec.~\ref{Sec: Conclusion}.


\section{Model and method}
\label{Sec: ModelMethod}

We consider the Hubbard Hamiltonian for spin-$\frac{1}{2}$ fermions on a square lattice with intersite hopping amplitudes $t_{jj^\prime}$ and a local repulsive interaction $U>0$
\begin{equation} \label{eq: HubbardHamiltonian}
 H = H_0 + H_{\text{int}} =
 \sum_{j,j',\sigma} t_{jj'} c^\dagger_{j\sigma} c^{\phantom\dagger}_{j'\sigma} +
 U \sum_j n_{j\uparrow} n_{j\downarrow} \, ,
\end{equation}
where $c_{j\sigma} (c^\dagger_{j\sigma})$ annihilates (creates) an electron with spin orientation $\sigma \in \{\uparrow,\downarrow\}$ on lattice site $j$ and 
$n_{j\sigma} = c^\dagger_{j\sigma} c_{j \sigma}$. The hopping matrix depends only on the relative distance between lattice sites $j$ and $j^\prime$. We choose $t_{jj^\prime} = -t$, if $j$ and $j^\prime$ are nearest-neighbor sites, $t_{jj^\prime} = -t^\prime$ if $j$ and $j^\prime$ are next-nearest-neighbor sites and $t_{jj^\prime} = 0$ otherwise.
Fourier transforming the hopping amplitudes yields the bare dispersion relation
\begin{equation}
 \eps_\bk = -2t (\cos k_x + \cos k_y) - 4t' \cos k_x \cos k_y \, .
\end{equation}
%


\subsection{Real-space mean-field Hamiltonian}
\label{Sec: MF Procedure}

We analyze the Hubbard model using a renormalized mean-field approach \cite{Wang2014, Yamase2016}, with an unbiased and unrestricted solution of the renormalized mean-field equations as in Ref.~\cite{Scholle2023}.
A direct Hartree-Fock decoupling of the Hubbard Hamiltonian would yield a mean-field Hamiltonian that does not capture superconductivity. We therefore use a functional renormalization group (fRG) flow to obtain a renormalized magnetic effective interaction $U^m$ and an effective non-local pairing interaction $U^p$ by integrating out high-energy degrees of freedom \cite{Wang2014}. We provide more details on the fRG method in Sec.~\ref{sec: fRG}.
The resulting mean-field Hamiltonian takes the form
\begin{equation} \label{eq: BdG_Hamiltonian}
 H^{\text{MF}} = H_0 + H_{\rm m}  + H_{\rm p} +\text{const}\, ,
\end{equation}
where the terms $H_{\rm m}$ and $H_{\rm p}$ represent the magnetic and pairing contributions, respectively. The latter can trigger superconductivity. The magnetic term, which may also lead to charge order, is given by
\begin{equation} \label{eq: Hm}
\begin{split}
 H_{\rm m} = & \sum_{j\sigma} \left(A_{j\bar{\sigma}} - \mu\right) n_{j\sigma} \\
 + & \sum_j \left( A_{j-} c^{\dagger}_{j\uparrow} c_{j\downarrow}
   + A_{j+} c^\dagger_{j\downarrow} c_{j\uparrow} \right) \, ,
\end{split}
\end{equation}
with the conventions $\bar{\uparrow} = \downarrow$ and $\bar{\downarrow} = \uparrow$.
The pairing term, including arbitrary spin singlet pairing between nearest neighbors on the lattice, has the form
\begin{equation} \label{eq: H sc}
\begin{split}
 H_{\rm sc} = \sum_j \big[
 & \Delta_j^x \left( c_{j\uparrow}^\dagger c_{j+\hat{x} \downarrow}^\dagger -
   c_{j\downarrow}^\dagger c_{j+\hat{x} \uparrow}^\dagger \right) \\
 + & \Delta_j^y \left( c_{j\uparrow}^\dagger c_{j+\hat{y}^\dagger \downarrow} -
   c_{j\downarrow}^\dagger c_{j+\hat{y} \uparrow}^\dagger \right) + \text{h.c.} \big] \, .
\end{split}
\end{equation}

The parameters $A_{j\sigma}$ and $A_{j\pm}$ are determined self-consistently through the following relations
\begin{subequations}\label{eq: DefineAs}
    \begin{align}
        &A_{j\sigma} = U^m \langle n_{j\sigma}\rangle,\\
        &A_{j+} = A_{j-}^* = -U^m \langle c_{j\uparrow}^\dagger c_{j\downarrow} \rangle.
    \end{align}
\end{subequations}
These parameters are directly related to the local charge and spin expectation values via
\begin{subequations} \label{eq: Relation Spins Charges Exp. Values}
 \begin{align}
  & \langle n_j \rangle = \frac{A_{j\uparrow}+A_{j\downarrow}}{U^m} \, , \\
  & \langle S^z_j \rangle = \frac{A_{j\uparrow}-A_{j\downarrow}}{2U^m} \, , \\
  & \langle S^x_j \rangle = -\frac{A_{j+}+A_{j-}}{2U^m} \, , \\
  & \langle S^y_j \rangle = -\frac{A_{j+}-A_{j-}}{2i U^m} \, ,
 \end{align}
\end{subequations}
where $n_j = n_{j\uparrow} + n_{j\downarrow}$ and $\vec{S}_j =\frac{1}{2} c_j^\dagger \vec{\sigma}c_j,$ with the Pauli matrices $\vec{\sigma} = (\sigma^x,\sigma^y,\sigma^z)$.

The pairing gaps are related to pairing expection values and the pairing interaction by
\begin{equation} \label{eq: SC Orderparameter Definition}
\begin{split}
 \Delta_j^x & = \frac{1}{2} U^p \left(
 \left\langle c_{j \uparrow} c_{j + \hat{x} \downarrow} \right\rangle -
 \left\langle c_{j \downarrow} c_{j + \hat{x} \uparrow} \right\rangle \right)
 \, , \\
 \Delta_j^y & = \frac{1}{2} U^p \left(
 \left\langle c_{j \uparrow} c_{j + \hat{y} \downarrow} \right\rangle -
 \left\langle c_{j \downarrow} c_{j + \hat{y} \uparrow} \right\rangle \right) \, .
\end{split}
\end{equation}
%
%
There are thus two complex superconducting order parameters per site, $\Delta_j^x$ and $\Delta_j^y$, corresponding to the adjacent bonds in $x$ and $y$ direction, respectively.
Homogeneous $d$-wave superconductivity and homogeneous extended $s$-wave superconductivity are recovered as the special cases where $\Delta_j^x = -\Delta_j^y = \Delta$ and $\Delta_j^x = \Delta_j^y = \Delta$, respectively.

If we set $U^p = 0$, which implies $\Delta^{x/y}_{j} = 0$, the mean-field Hamiltonian in Eq.~\eqref{eq: BdG_Hamiltonian} reduces to a conventional Hartree-Fock decoupling of the Hubbard Hamiltonian with interaction strength $U^m$~\cite{Scholle2023}.

Note that Eq. \eqref{eq: SC Orderparameter Definition} accounts exclusively for the singlet pairing component. While triplet pairing can theoretically emerge in states coexisting with magnetic order, its contribution has been found to be negligible relative to the singlet component in this regime.

The total renormalized mean-field Hamiltonian can be written in a compact matrix form as
\begin{equation}\label{eq: BdG_Ham Matrix Form}
    H^{\text{MF}} = \frac{1}{2} \,
    \sum_{j,j'} \sum_{\sigma,\sigma'} \sum_{\alpha,\alpha'}
    d^\dagger_{j\sigma\alpha} \mathcal{H}_{jj'}^{\sigma\sigma'\alpha\alpha'}
    d_{j'\sigma'\alpha'} + \text{const} \, ,
\end{equation}
where the index $\alpha \in \{+,-\}$ distinguishes between normal ($d_{j\sigma +} = c_{j\sigma}$) and anomalous ($d_{j\sigma -} = c^\dagger_{j\sigma}$) operators.
The matrix $\mathcal{H}_{jj'}^{\sigma\sigma'\alpha\alpha'} \in
\mathbb{C}^{4\mathcal{N} \times 4\mathcal{N}}$ depends on $8\mathcal{N}$ real parameters that must be determined self-consistently.

To calculate the expectation values from Eqs.~\eqref{eq: Relation Spins Charges Exp. Values} and \eqref{eq: SC Orderparameter Definition}, we first define the matrix
\begin{equation} \label{eq: MMatrix T!=0}
 \mathcal{M}_{jj'}^{\sigma\sigma'\alpha\alpha'} =
 \sum_{\ell=1}^{4\mathcal{N}} (v^\ell_{j\sigma\alpha})^* v^\ell_{j'\sigma'\alpha'}
 f(\eps_\ell) \, ,
\end{equation}
where $f(x) = (e^{x/T}+1)^{-1}$ is the Fermi function, and $\ell$ labels the $4\mathcal{N}$ eigenvalues $\epsilon_\ell$ and corresponding eigenvectors $v_{j\sigma\alpha}^\ell$ of $\mathcal{H}$.

From this matrix, we obtain the charge, spin, and superconducting expectation values as
\begin{equation}\label{eq: Expectation values from M matrix}
    \begin{split}
        \langle n_{j\sigma} \rangle &= \frac{1}{2} \left( \mathcal{M}_{jj}^{\sigma\sigma++} - \mathcal{M}_{jj}^{\sigma\sigma--} + 1\right),\\
        \langle c_{j\uparrow}^\dagger c_{j\downarrow} \rangle &= \frac{1}{4} \big( \mathcal{M}_{jj}^{\uparrow\downarrow++} + {\mathcal{M}^{\downarrow\uparrow++}_{jj}}^* \\
        & \qquad - {\mathcal{M}_{jj}^{\uparrow\downarrow--}}^* -{\mathcal{M}_{jj}^{\downarrow\uparrow--}} \big),\\
        \left \langle c_{j \uparrow}c_{j' \downarrow} \right \rangle &= \frac{1}{4} \big( {\mathcal{M}_{jj^\prime}^{\uparrow\downarrow-+}} - {\mathcal{M}_{j^\prime j}^{\downarrow\uparrow-+}}\\
        &\qquad +
        {\mathcal{M}_{j^\prime j}^{\downarrow\uparrow+-}}^* -
        {\mathcal{M}_{j j^\prime}^{\uparrow\downarrow+-}}^* \big).
    \end{split}
\end{equation}
We solve this system self-consistently by an iterative procedure. We begin with a random initial set of parameters $\{ A_{j\sigma}, A_{j\pm}, \Delta_j^x,\Delta_j^y \}$, which are used to construct the Hamiltonian matrix $\mathcal{H}$. We compute its eigenvectors and eigenvalues and update the expectation values using Equation~\eqref{eq: Expectation values from M matrix}. This process is repeated until the expectation values converge.

Since we work at a fixed particle density $n$ rather than a fixed chemical potential, we adjust the chemical potential in each iteration, as detailed in Appendix~\ref{App: Numerical Details}. The free energy is given by
\begin{eqnarray}\label{eq: Free energy}
 F &=& - \frac{T}{2} \sum_\ell  \ln(1 + e^{-\beta \epsilon_\ell })
 + (\mu + U^m/2) \sum_j \langle n_j \rangle - \mu\mathcal{N} \nonumber \\
 && - \frac{1}{U^m} \sum_j \left( A_{j\uparrow} A_{j\downarrow} - A_{j+} A_{j-} \right)
 \nonumber \\
 && + \frac{2}{U^p} \sum_j \left(\Delta_j^x \Delta_j^{x*} +
  \Delta_j^y \Delta_j^{y*} \right) \, .
\end{eqnarray}
%


\subsection{Classification of magnetic states}
\label{Sec: Classification}

In this section, we introduce a classification scheme for the obtained magnetic patterns. We classify the different states based on their spin pattern, following the approaches of Refs.~\cite{Sachdev2019, Scholle2023}.
In line with Ref.~\cite{Sachdev2019}, we generally find that the modulation of the charge density is approximately proportional to the square of the spin amplitudes, that is,
\begin{equation}\label{eq: charge equal spin^2}
 \langle n_j \rangle \approx
 \mbox{const} \, \langle \Vec{S}_j \rangle^2 + \mbox{const} \, ,
\end{equation}
as can be deduced from the lowest order coupling between the spin and charge density wave order parameters in a Landau theory~\cite{ZacharKivelson1998}.

To classify a magnetic state defined by the spin expectation values $\langle \Vec{S}_j \rangle$, we start from the Fourier representation,
\begin{equation}\label{Classification_state_ala_Sachdev}
 \big\langle \Vec{S}_j \big\rangle =
 \sum_{\bq} \Vec{\mathcal{S}}_{\bq}\, e^{i\bq\cdot \mathbf{r}_j}\,,
\end{equation}
where $\Vec{\mathcal{S}}_{\bq}$ is the Fourier component of the spins corresponding to the wave vector $\bq$, and $\mathbf{r}_j$ is the real-space coordinate of lattice site $j$. Since the spin expectation values are real-valued, its Fourier components obey the relation
$\vec{\mathcal{S}}_\bq = \vec{\mathcal{S}}^*_{-\bq}$.
The sum runs over the $\mathcal{N} = \mathcal{N}_x \mathcal{N}_y$ momenta allowed by the periodic boundary conditions of the finite-size system. Since we always choose $\mathcal{N}_x$ and $\mathcal{N}_y$ as even numbers, the allowed momenta can be written as
\begin{equation}\label{eq: Allowed momenta}
 \bq \in \left\{ \left. \left(
 \pi - 2\pi\frac{\nu_x}{\mathcal{N}_x},\pi - 2\pi\frac{\nu_y}{\mathcal{N}_y} \right)
 \right| \nu_\alpha \in \left\{ 0,1,\dots,\mathcal{N}_\alpha-1\right\} \right\} .
\end{equation}
In the majority of cases, only a single mode with a fixed wave vector $\bq$ (together with its partner $-\bq$) contributes significantly to the spin pattern $\langle \vec{S}_j \rangle$. This dominant wave vector typically takes the form
\begin{equation}
 \bQ_x \equiv (\pi - 2\pi\eta_x, \pi) \quad \text{or} \quad
 \bQ_y \equiv (\pi, \pi - 2\pi\eta_y) \, ,
\end{equation}
where the parameters $\eta_\alpha$ are commonly referred to as ``incommensurabilities''.
In a finite system, $\eta_\alpha$ is restricted to integer multiples of $1/\mathcal{N}_\alpha$ (see Eq.~\eqref{eq: Allowed momenta}).
We also checked for magnetic order with diagonal wave vectors of the form $\bQ_{xy} \equiv (\pi-2\pi\eta,\pi-2\pi\eta)$, but wave vectors of this kind do not occur in any of the states we obtained.

We can therefore express $\vec{\mathcal{S}}_\bq$ as
\begin{equation} \label{eq: Form of Sq}
 \vec{\mathcal{S}}_\bq = \frac{1}{2} 
 \left( \vec{\mathcal{S}} \, \delta_{\bq,\bQ} +
 \vec{\mathcal{S}}^* \, \delta_{\bq,-\bQ} \right) , 
\end{equation}
where $\bQ$ has the form $\bQ_x$ or $\bQ_y$. 
We distinguish between five different phases. Configurations that can be transformed into each other by a point group symmetry operation of the lattice (rotations and reflections), a spatial translation (changing the phase of $\vec{\mathcal{S}}$), or by a global SU(2) rotation of the spin frame, belong to the same phase. In the following, we describe each phase using a representative spin configuration.

In principle, also other spin configurations are possible, for example the phases $F$ and $G$ in the classification of Ref.~\cite{Sachdev2019}, as well as the coplanar or collinear bidirectional stripes of Refs.~\cite{Sachdev2019,Scholle2023}, but we do not find them in the parameter regime explored in our calculations.


\subsubsection{Paramagnetic}

In the paramagnetic phase, the expectation value of the spin on each site vanishes, that is,
$\langle \Vec{S}_j \rangle = 0$, and the charge distribution is uniform, $\langle n_j \rangle = n$.


\subsubsection{N\'eel antiferromagnetism}

In a N\'eel antiferromagnet, only $\vec{\mathcal{S}}_{(\pi,\pi)}$ is nonzero. In real space, adjacent spins are collinear and point in opposite directions. All spins have the same amplitude, and the charge distribution is homogeneous.


\subsubsection{Spiral order}

A magnetic spiral state is characterized by a single Fourier mode with a wave vector $\bQ$, and (as a representative) 
\begin{equation}\label{eq: Definition Spiral State}
 \Vec{\mathcal{S}} = \mathcal{S}_0 \begin{pmatrix} 1 \\ i \\ 0 \end{pmatrix} ,
\end{equation}
with a real $\mathcal{S}_0$.
In real space, the corresponding spin pattern takes the form 
\begin{equation}
 \big\langle \Vec{S}_j \big\rangle = \mathcal{S}_0
 \begin{pmatrix}
 \cos(\bQ \cdot \mathbf{r}_j) \\
 \sin(\bQ \cdot\mathbf{r}_j) \\
 0
 \end{pmatrix}.
\end{equation}
The spin amplitude $|\langle \Vec{S}_j \rangle|$ is constant (independent of $j$), and thus there is no charge modulation in this state. The wave vector $\bQ$ has the form $\bQ_x = (\pi-2\pi\eta,\pi)$ or $\bQ_y = (\pi,\pi-2\pi\eta)$, that is, only one component differs from $\pi$. The N\'eel state is the special case of a spiral with $\bQ = (\pi,\pi)$.


\subsubsection{Stripe order}

A (unidirectional) stripe state is a spin pattern described by a single $\bq$-mode such that 
\begin{equation}
 \vec{\mathcal{S}} = \mathcal{S}_0 \begin{pmatrix} 1 \\ 0 \\ 0 \end{pmatrix},
\end{equation}
and the corresponding real-space representation
\begin{equation}\label{eq: PERFECT STRIPES}
 \big\langle \Vec{S}_j \big\rangle = \mathcal{S}_0
 \begin{pmatrix}
 \cos(\bQ \cdot \mathbf{r}_j) \\ 0 \\ 0 \end{pmatrix}.
\end{equation}
As in the spiral state, the wave vector $\bQ$ takes either the form $\bQ_x$ or $\bQ_y$.
The spins are ordered collinearly with their amplitude modulated along the $x$- or $y$-axis. 

In this state, the charge order is a unidirectional charge density wave with wave vector $2\bQ$, that is, $\langle n_j \rangle - n \propto \cos(2\bQ \cdot \mathbf{r}_j)$. The minima of the charge modulation coincide with the minima in the spin amplitude.
For $\bQ = (\pi,\pi)$ one again recovers the N\'eel state as a special case.


\subsubsection{Strange orders}

Some magnetic states do not fit into any of the previously described categories.
Such states do not have a simple spin structure in momentum space and often exhibit complex charge patterns. Throughout this work, we will refer to those orders as ``strange order''.
In the thermodynamic limit, these states most likely disappear, as shown in Ref.~\cite{Scholle2023}.


\subsection{Functional RG and effective interactions} \label{sec: fRG}

In this section, we detail the calculation of the effective magnetic and pairing interactions, $U^m$ and $U^p$, for the mean-field Hamiltonian in Eq.~\eqref{eq: BdG_Hamiltonian}.
These interactions are derived using the functional renormalization group (fRG) \cite{Metzner2012Review}, which integrates out high-energy degrees of freedom to obtain effective low-energy interactions. Specifically, we use the temperature flow approach \cite{Honerkamp2001} in an approximation used previously in Ref.~\cite{Bonetti2022gauge}.

Temperature can be used as a flow parameter after rescaling the fermionic field as $\psi_j \mapsto T^{3/4} \psi_j$, and defining a rescaled bare Green's function
$G^T_0(\bk,\nu_n) = T^{1/2}/(i\nu_n + \mu - \epsilon_\bk)$,
where $\nu_n = (2n+1) \pi T$ with integer $n$ is the fermionic Matsubara frequency.
The flow proceeds continuously from high to low temperatures, and thus involves a successive integration of degrees of freedom from high to low energies \cite{Honerkamp2001}.

We approximate the fRG flow by a second order (one-loop) flow of the two particle vertex $V^T$, discarding self-energy feedback and contributions from the three-particle vertex.
In the symmetric regime of the flow, the vertex has the spin SU(2) invariant structure
\begin{equation}
\begin{split}
 V_{\sigma_1,\sigma_2,\sigma_3,\sigma_4}^T(k_1,k_2,k_3,k_4) & =
 V^T(k_1,k_2,k_3,k_4) \delta_{\sigma_1\sigma_3} \delta_{\sigma_2\sigma_4} \\
 & - V^T(k_2,k_1,k_3,k_4) \delta_{\sigma_1\sigma_4} \delta_{\sigma_2\sigma_3} \, ,
\end{split} \nonumber
\end{equation}
where $V^T(k_1,k_2,k_3,k_4) = V_{\uparrow\downarrow\uparrow\downarrow}^T(k_1,k_2,k_3,k_4)$, and the variables $k_i$ are combined momentum and frequency variables.
We discard the frequency dependence of the vertex function, which is irrelevant in power-counting. To parametrize the momentum dependence, we use the channel decomposition \cite{Husemann2009}
\begin{equation} \label{eq: vertex_function}
\begin{split}
 & V^T(\bk_1, \bk_2, \bk_3, \bk_4) = U \\
 & + \phi^{m,T}_{\frac{\bk_1 + \bk_4}{2}, \frac{\bk_2 + \bk_3}{2}}(\bk_2-\bk_3)
 + \frac{1}{2}\phi^{m,T}_{\frac{\bk_1 + \bk_3}{2}, \frac{\bk_2 + \bk_4}{2}}(\bk_3-\bk_1) \\
 & - \frac{1}{2}\phi^{c,T}_{\frac{\bk_1 + \bk_3}{2}, \frac{\bk_2 + \bk_4}{2}}(\bk_3-\bk_1)
 - \phi^{p,T}_{\frac{\bk_1 - \bk_2}{2}, \frac{\bk_3 - \bk_4}{2}}(\bk_1 + \bk_2) \, ,
\end{split}
\end{equation}
where the functions $\phi^{m,T}$, $\phi^{c,T}$, and $\phi^{p,T}$ capture fluctuation contributions in the magnetic, charge, and pairing channels, respectively.
The dependencies of these functions on the linear combination of momenta in brackets are much stronger than those in the subscripts. Hence, we represent the latter dependencies by a small number of form factors \cite{Husemann2009}.

We run the fRG flow from $T = \infty$ down to the critical temperature $T^*$ at which $V^T$ diverges. In a broad region around half filling, the divergence occurs in the magnetic channel, and in the $d$-wave pairing channel at larger electron or hole doping.
To enter the symmetry-broken regime at temperatures below $T^*$, we simplify the flow by decoupling the various channels, and allow for magnetic and superconducting order parameters in the form of anomalous self-energy contributions. One can show that this simplified flow can be formally integrated, yielding mean-field equations for the order parameters with renormalized effective interactions \cite{Wang2014}. These renormalized interactions are the two-particle irreducible parts of the full vertex $V^T$ in the symmetry-breaking channels, and can be computed from the full vertex by solving a Bethe-Salpeter equation.

Specifically, in the magnetic channel, the Bethe-Salpeter equation takes the form
\begin{eqnarray} \label{eq: BS_mag}
 V^{m,T^*}_{\bk,\bk'}(\bq) &=& \bar{V}^{m}_{\bk,\bk'}(\bq) \nonumber \\
 &-& \int_{\bk''} \bar{V}^{m}_{\bk,\bk''}(\bq) \Pi_{\bk''}^{ph,T^*}(\bq)
 V^{m,T^*}_{\bk'',\bk'}(\bq) \, ,
\end{eqnarray}
with $V^{m,T}_{\bk,\bk'}(\bq) = V^T(\bk - \bq/2,\bk'+\bq/2,\bk'-\bq/2,\bk+\bq/2)$,
and in the pairing channel,
\begin{eqnarray} \label{eq: BS_pair}
 V^{p,T^*}_{\bk,\bk'}(\bq) &=& \bar{V}^{p}_{\bk,\bk'}(\bq) \nonumber \\
 &-& \int_{\bk''} \bar{V}^{p}_{\bk,\bk''}(\bq) \Pi_{\bk''}^{pp,T^*}(\bq)
 V^{p,T^*}_{\bk'',\bk'}(\bq),
\end{eqnarray}
where $V^{p,T}_{\bk,\bk'}(\bq) = V^T(\bq/2-\bk,\bq/2+\bk,\bq/2-\bk',\bq/2+\bk')$.
Here, $\bar{V}^m$ and $\bar{V}^p$ denote the irreducible vertices, and $\Pi^{ph}$ and $\Pi^{pp}$ are the particle-hole and particle-particle bubbles, respectively. These bubbles are calculated using the bare Green's functions at temperature $T^*$,
\begin{eqnarray} \label{eq: pi}
 \Pi_{\bk}^{ph,T^*}(\bq) &=&
 \sum_{\nu_n} G^{T^*}_0(\bk - \bq/2,i\nu_n) G^{T^*}_0(\bk + \bq/2,i\nu_n) \, ,
\nonumber \\
 \Pi_{\bk}^{pp,T^*}(\bq) &=&
 \sum_{\nu_n} G^{T^*}_0(\bq/2-\bk,-i\nu_n) G^{T^*}_0(\bq/2 + \bk,i\nu_n).
 \nonumber
\end{eqnarray}
%
The final effective interactions used in Eq.~\eqref{eq: BdG_Hamiltonian} are then obtained by a form-factor projection of the irreducible vertices:
\begin{equation}
\label{eq: Um Up}
\begin{split}
 U^m &= \int_{\bk,\bk'} \bar{V}^{m}_{\bk,\bk'}(\bq=\bQ) \, , \\
 U^p &= \int_{\bk,\bk'} d_\bk d_{\bk'}\bar{V}^{p}_{\bk,\bk'}(\bq=\bzero) \, ,
\end{split}
 \end{equation}
where $d_\bk = \cos k_x - \cos k_y$ is the $d$-wave pairing form factor.
The momentum $\bQ$ is chosen where the magnetic $s$-wave interaction $\int_{\bk,\bk'}V^{m,T^*}_{\bk,\bk'}(\bq)$ reaches its maximum. 

It should be noted that while $U^p$ is extracted specifically from the d-wave projection of the effective interaction, the real-space functional in Eq. \eqref{eq: SC Orderparameter Definition} allows for a general nearest-neighbor pairing. Consequently, the self-consistent solution is not restricted to $d$-wave symmetry and may, in principle, accommodate extended $s$-wave components and inhomogeneous states.


\section{Numerical results} \label{Sec: NumericalResults}

We now present our numerical results for the phase diagram of the Hubbard model. We chose a bare interaction strength of $U = 4t$, which is strong enough to obtain an extended regime of magnetic and superconducting states, while still remaining within the applicability range of the fRG+MF method. We set the next-to-nearest neighbor hopping to $t' = -0.2t$, which lies in the range of $t'$ values applying to cuprates.

\begin{figure}
\centering
 \includegraphics[width=0.4\textwidth]{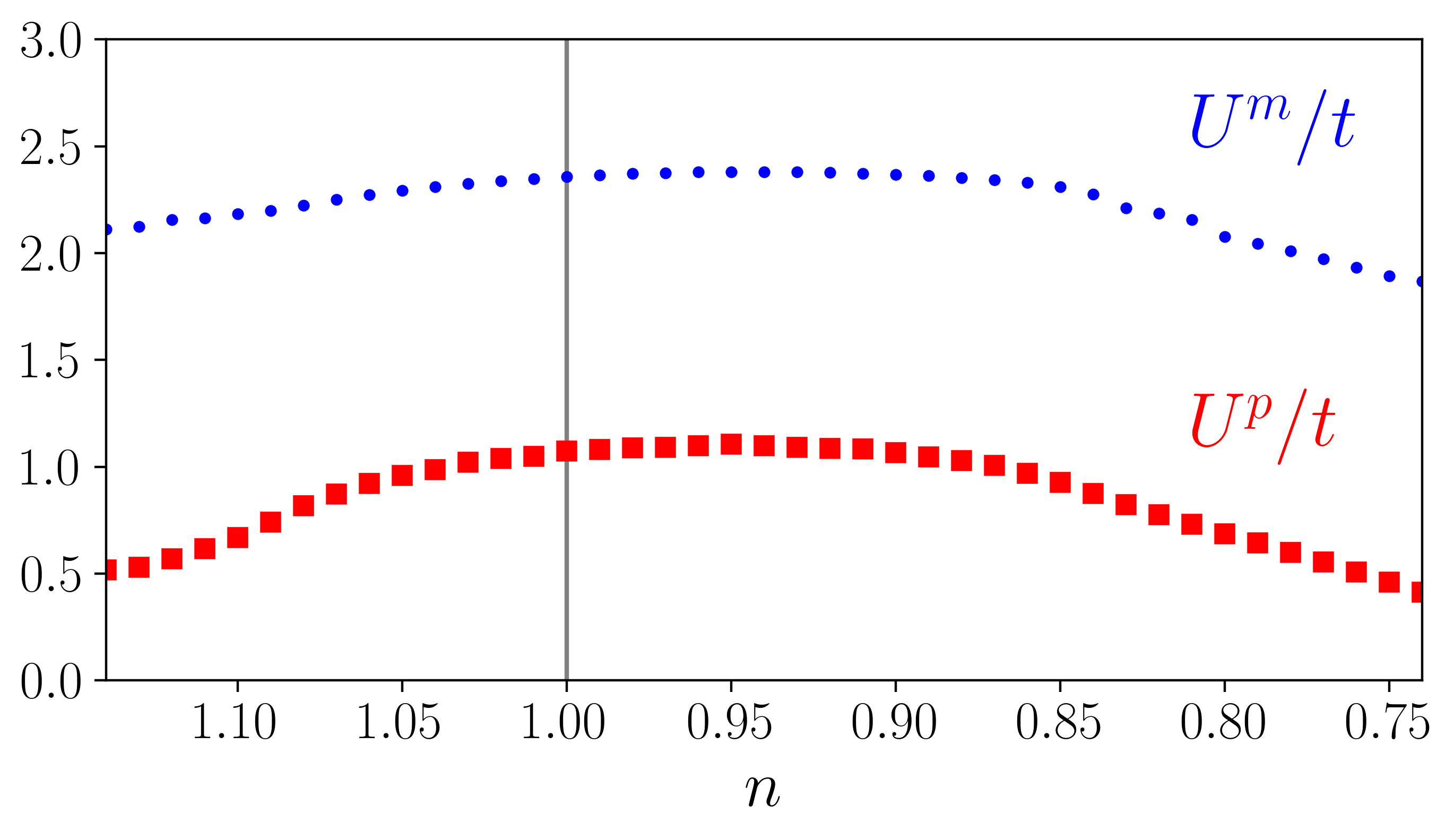}
 \caption{Effective interactions $U^m$ and $U^p$ obtained from the fRG flow as a function of filling $n$, for a bare interaction $U=4t$, and $t'=-0.2t$.}
\label{fig: renormalized Interactions}
\end{figure}
In Fig.~\ref{fig: renormalized Interactions}, we show the renormalized magnetic and pairing interactions, $U^m$ and $U^p$, we obtained from the fRG flow, as discussed in Section~\ref{sec: fRG}. The effective magnetic interaction $U^m$ ranges between $2t$ and $2.5t$, while the pairing interaction $U^P$ varies between $0.5t$ and $t$. Both interactions decrease in magnitude for large electron or hole doping.

We use these effective interactions to build a density-temperature phase diagram in steps of $0.01$ for both $n$ and $T/t$. To avoid trapping in local minima above the global minimum of the free energy, we evaluated each $(n,T)$ point of the phase diagram starting from a random configuration of expectation values twice, and then from the converged solution of neighboring points, that is, $(n \pm 0.01,T)$ and $(n, T \pm 0.01t)$. In most cases, the self-consistency loop converges to the same final state regardless of the initial choice. In the few cases where the various calculations converge to different states, we retain the state with the lowest free energy.


\subsection{Phase diagram} 

We now present the phase diagram as a function of the filling $n$ and temperature $T$ as obtained from calculations on a $28 \times 28$ lattice. We chose $\mathcal{N}_x = \mathcal{N}_y$ to avoid any bias in favor of anisotropic magnetic order.
\begin{figure}
\centering
 \includegraphics[width=0.48\textwidth]{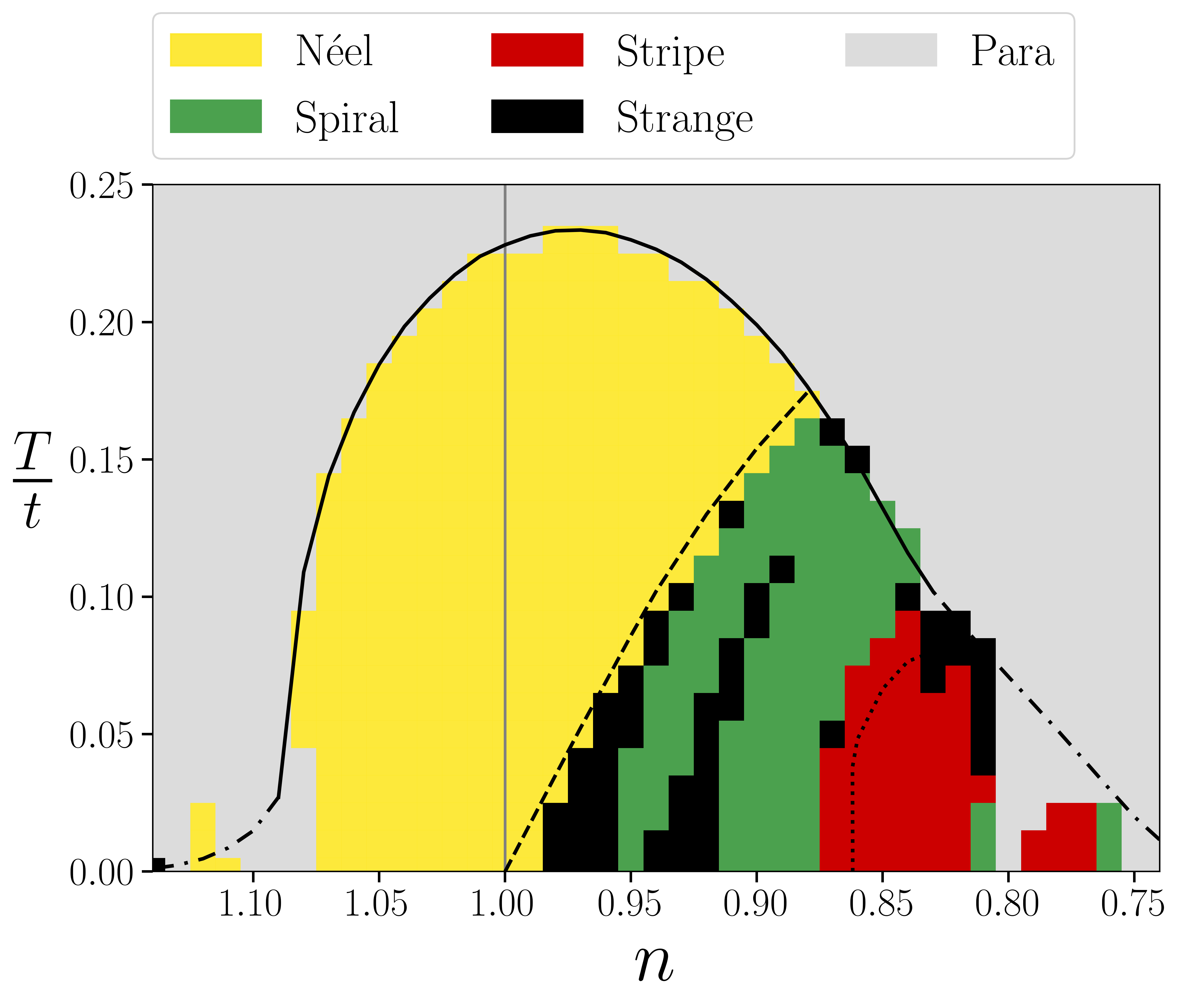}
 \caption{Magnetic phase diagram for $U = 4t$ and $t' = -0.2t$. The colored pixels show the results of the real-space calculation on a $28\times 28$ lattice. The black lines have been obtained in the thermodynamic limit. The solid black line shows the divergence of the fRG flow and represents the onset of the magnetically ordered phase. Along the black dash-dotted lines, the fRG flow diverges in the pairing channel, signaling the onset of superconductivity as the leading instability. The dashed line inside the magnetically ordered regime indicates the transition from N\'eel order to spiral magnetic order. The dotted line indicates an instability of the spiral phase revealed by a divergence of the charge susceptibility.}
\label{fig: MagneticPhaseDiagram}
\end{figure}
In Fig.~\ref{fig: MagneticPhaseDiagram} we show the various magnetic states, which partially coexist with superconductivity. The superconducting gap amplitude will be shown in a separate plot below.
The colored pixels represent the states calculated on the finite-size lattice, and the black envelope is the line corresponding to the critical temperature $T^*$ in the fRG flow. If the flow diverges in the magnetic channel, we draw $T^*$ as a solid line, which is the case for $0.83 \lesssim n \lesssim 1.09$. If it diverges in the pairing channel, we indicate the divergence by a dash-dotted line.
By construction, the mean-field equations yield a symmetric phase for $T>T^*$ and a symmetry-broken phase for $T<T^*$. The solid and dash-dotted black lines coincide well with the onset of magnetic or superconducting order obtained from the finite-size calculations, respectively. In this respect the $28 \times 28$ lattice is already a good approximation to the thermodynamic limit. We identify the converged spin patterns obtained from our real-space Hartree-Fock calculations according to the classification in Sec.~\ref{Sec: Classification}.

We find a N\'eel ordered dome around half-filling, with a maximum critical temperature $T^* \approx 0.23t$ in the slightly hole-doped regime, at $n \approx 0.97$.
In the electron-doped part of the phase diagram, we exclusively find N\'eel antiferromagnetism.

As we increase the hole doping, a region of spiral states emerges. These spiral states exhibit an incommensurability of $\eta = \frac{1}{28}$ (close to the N\'eel ordered regime) or $\eta = \frac{2}{28}$ for larger dopings. Between these two spiral regimes, as well as between the N\'eel and spiral orders, we find small regions of strange states. In these regions, it is energetically unfavorable for the system to order with one of the allowed spiral momenta from Eq.~\eqref{eq: Allowed momenta}. Since intermediate momenta are forbidden by the periodic boundary conditions, the system orders in these strange patterns, where the dominant wave vectors correspond to the incommensurabilities of the adjacent spiral (or N\'eel) regimes. Analogous to the findings in Ref.~\cite{Scholle2023}, we consider these strange ordered phases to be finite-size artifacts that disappear in the thermodynamic limit. Thus, the real-space data are consistent with a spiral state in the thermodynamic limit, with a smoothly varying, generally incommensurate $\eta$ which increases with increasing hole doping or decreasing temperature.

Directly at the critical temperature $T^*$, we can additionally determine the type of magnetic order that emerges in the thermodynamic limit, based on a Landau theory \cite{Scholle2024}, which is at $T = T^*$ equivalent to the mean-field procedure we described in Sec.~\ref{Sec: MF Procedure}. We performed these calculations for the dopings where a magnetic instability in the fRG flow occurs, that is, along the solid black line in Fig.~\ref{fig: MagneticPhaseDiagram}. We confirm N\'eel order for the entire electron-doped side, as well as for fillings down to $n \approx 0.88$. For lower fillings, we find that the spiral order is the leading instability, in agreement with our real-space data.

For low temperatures, the spiral regime extends up to a hole doping of $n = 0.88$. This represents a much wider interval than that obtained from a pure Hartree-Fock analysis \cite{Scholle2023} for a comparable $t^\prime$. We believe this extended range is due to the reduced magnetic interaction strength and the coexistence with superconductivity, which both favor the spiral order. For even higher hole doping values, we observe an extended region of stripe order around $n \approx 0.85$. These stripe states have an incommensurability $\eta = \frac{3}{28}$.

The dashed line in Fig.~\ref{fig: MagneticPhaseDiagram} shows the N\'eel-spiral transition obtained directly in the thermodynamic limit (analogous to Ref.~\cite{Scholle2023}), which matches our finite-size results remarkably well.
The dotted line marks the divergence of the charge susceptibility in the spiral state (see Sec.~\ref{Sec: Susceptibilities}), signaling the onset of more complex orders that eventually evolve into stripe states \cite{Scholle2024}.

\begin{figure}
\centering
 \includegraphics[width=0.48\textwidth]{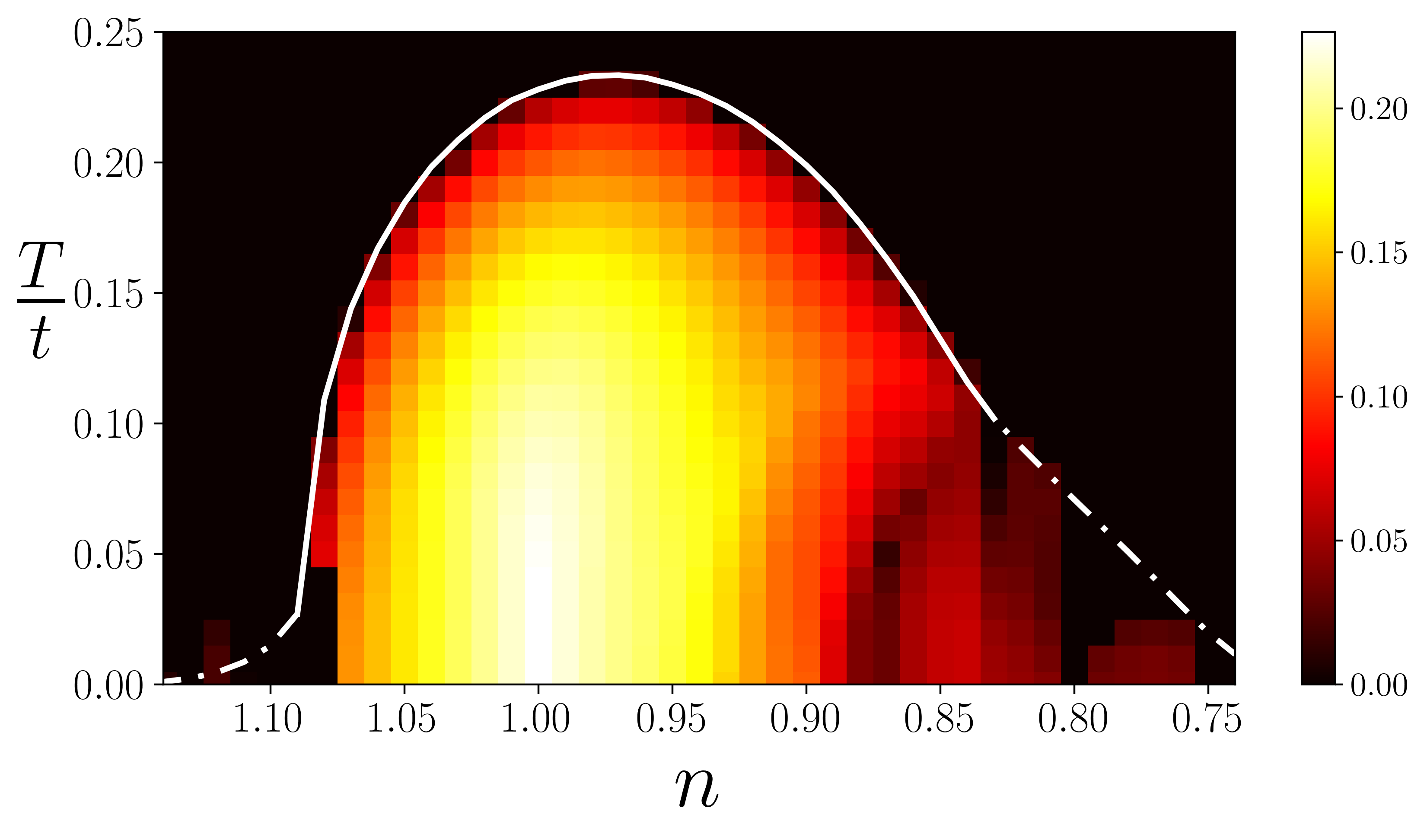}
 \caption{Average magnetization $m$ as a function of filling $n$ and temperature $T$ for the parameters from Fig.~\ref{fig: MagneticPhaseDiagram}. The white line represents $T^*$. The magnetization is highest for low temperatures near half-filling.}
\label{fig: Magnetization}
\end{figure}

In Fig.~\ref{fig: Magnetization}, we show the average magnetization
\begin{equation}
 m = \frac{1}{\mathcal{N}}\sum_j
 \sqrt{\langle S^x_j\rangle^2 + \langle S^y_j\rangle^2 + \langle S^z_j\rangle^2} \,
\end{equation}
as a function of $n$ and $T$.
Within the resolution of our method, the transition from the paramagnetic state to the magnetically ordered state appears to be a second-order transition. The strongest magnetization appears at half-filling and for low temperatures.
In the region where the critical temperatures are due to a magnetic instability in the fRG calculations (that is, the solid white line in Fig.~\ref{fig: Magnetization}), the magnetization persists up to the critical temperature, while for the regions where pairing is the leading instability (dash-dotted lines), the magnetization vanishes before reaching $T_c$.

The magnetization weakens with increasing hole doping, with $m$ reaching zero at $n = 0.8$, before a small region of weak magnetization reemerges for $0.75 < n <0.8$. In this region, we find mostly stripe order with an incommensurability of $\eta = \frac{4}{28}$.
For high electron or hole dopings ($n > 1.08$ or $n < 0.75$), we find paramagnetic states, in which the spin expectation values vanish on each site, while the superconducting order parameter remains finite.
In Sec.~\ref{sec: Thermodyn Limit} we show that in the thermodynamic limit, the magnetization disappears at van Hove filling and reemerges on both sides in doping. To show that, we recalculated the unrestricted HF phase diagram in this region on $40 \times 40$ lattices, as well as in the thermodynamic limit restricted to spiral order.

\begin{figure}
\centering
 \includegraphics[width=0.48\textwidth]{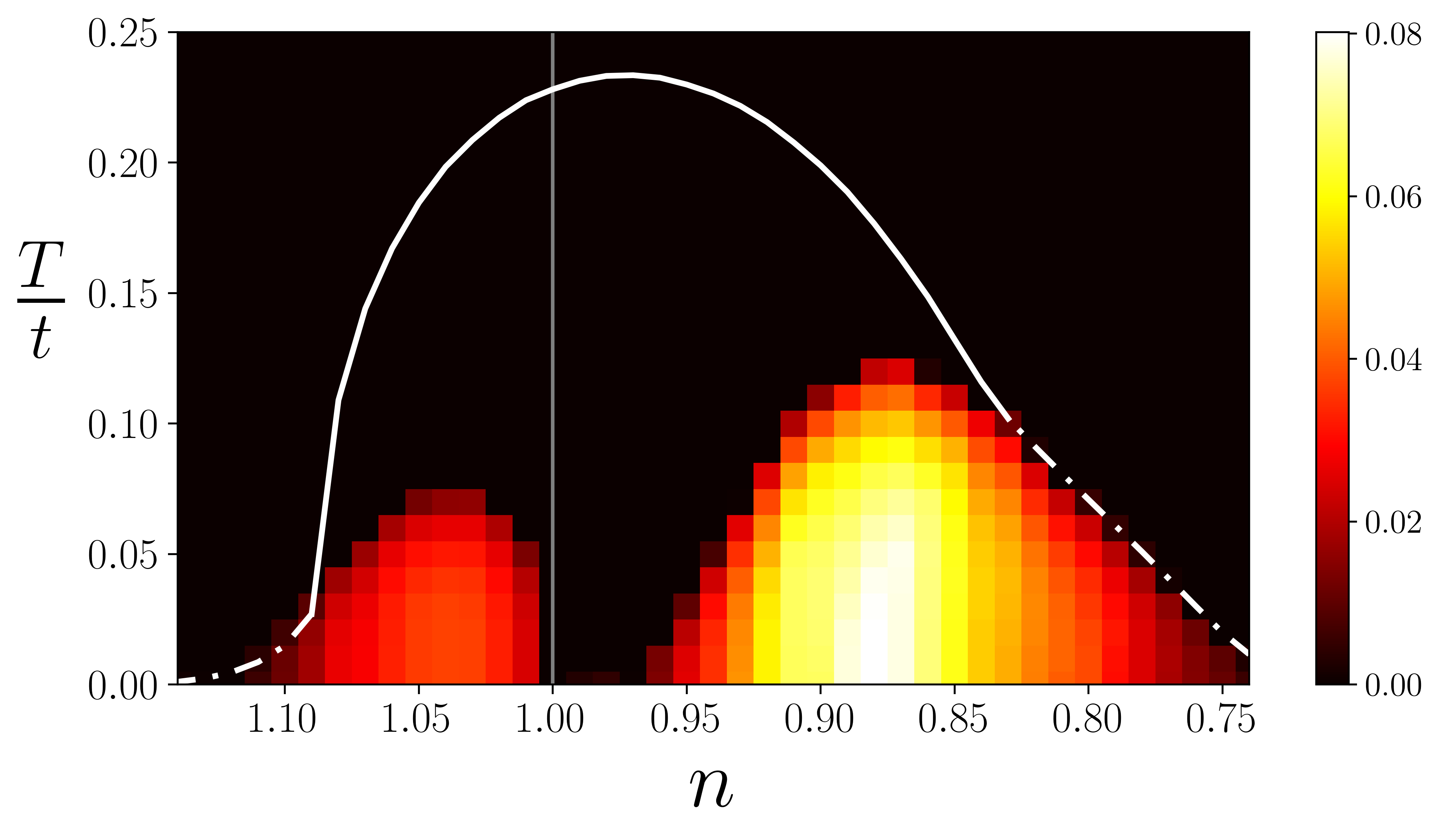}
 \caption{Average superconducting amplitude $\Delta$ in units of $t$ as a function of filling $n$ and temperature $T$ for the parameters from Fig.~\ref{fig: MagneticPhaseDiagram}. The white line represents $T^*$.}
\label{fig: Superconducting amplitude}
\end{figure}

In Fig.~\ref{fig: Superconducting amplitude}, we plot the average superconducting amplitude $\Delta = \frac{1}{\mathcal N\sqrt{2}} \sum_j\sqrt{|{\Delta_j^x}|^2 + |{\Delta_j^y}|^2}$ as a function of filling and temperature.
In the electron-doped regime, the superconducting critical temperature rises steeply when moving away from half filling, reaching its maximum value $T_c = 0.07t$ at $n \approx 1.04$. The superconducting amplitude in this regime remains relatively small, $\Delta < 0.05t$. On the hole-doped side, the superconducting critical temperature rises more slowly than on the electron-doped side. However, the maximal superconducting amplitude and the highest critical temperature are both larger than on the electron-doped side with the average superconducting amplitude reaching up to $\Delta = 0.11t$ and the critical temperature peaking at $T_c = 0.12t$ around $n \approx 0.87$. In both regimes, the superconducting state always has $d_{x^2-y^2}$-wave symmetry.

In a large region of the phase diagram, magnetic and superconducting orders coexist. While the N\'eel and spiral phases occur with as well as without superconductivity, the stripe phase always coexists with superconductivity for the set of parameters we used.

\begin{figure*}
\centering
 \includegraphics[width=0.8\textwidth]{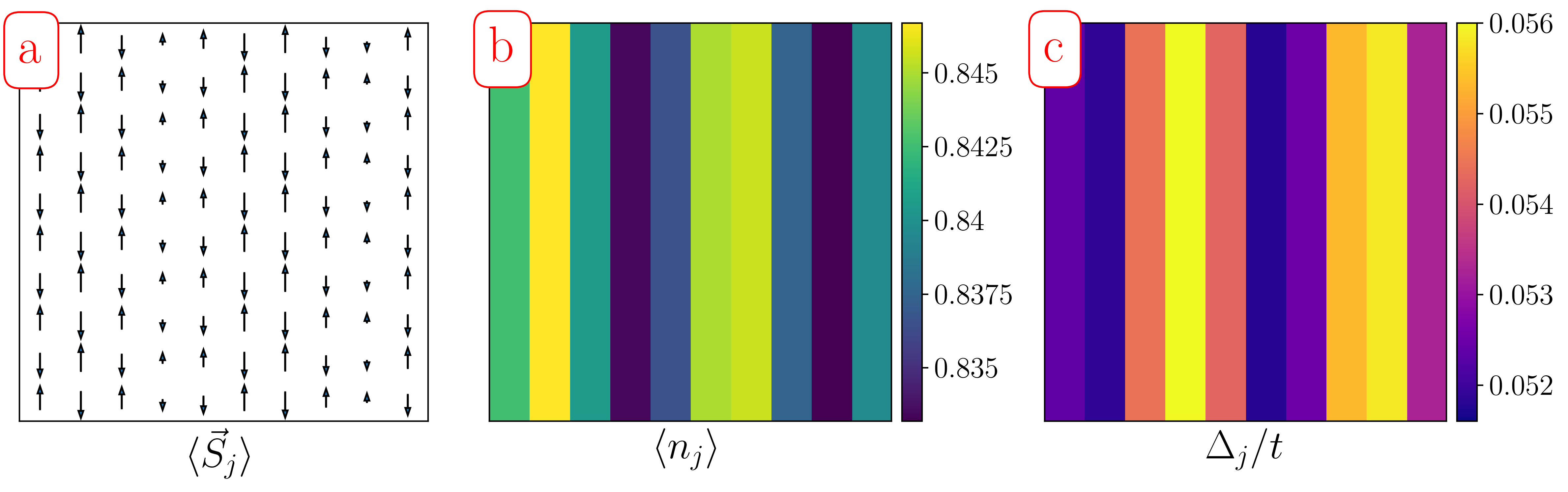}
 \caption{Magnetic, charge, and superconducting order pattern for the stripe state we find at $T = 0.02t$, $n = 0.84$. Only $10 \times 10$ sites are shown here for better visibility. In panel (a) one can see the collinear spin pattern with its amplitude modulated in $x$-direction. In panel (b) we see the charge pattern, which is also modulated along the $x$-axis with the lowest filling coinciding with the smallest local magnetization. In panel (c) the superconducting amplitude $\Delta_j$ on each site is shown. It is also modulated along the $x$-axis with the superconducting amplitude being maximal where the local filling is minimal.}
\label{fig: 3 Stripe patterns}
\end{figure*}

In the stripe phase, we observe a spatial modulation of the superconducting amplitude,
$\Delta_j = \frac{1}{\sqrt{2}}( |{\Delta_j^x}|^2 + |{\Delta_j^y}|^2 )^{1/2}$,
as shown in Fig.~\ref{fig: 3 Stripe patterns}.
The modulation in superconductivity has the same wavelength as the modulation of the charge order with the maxima of the superconducting amplitude being located at the minima of the local fillings.

\begin{figure}
\centering
 \includegraphics[width=0.49\textwidth]{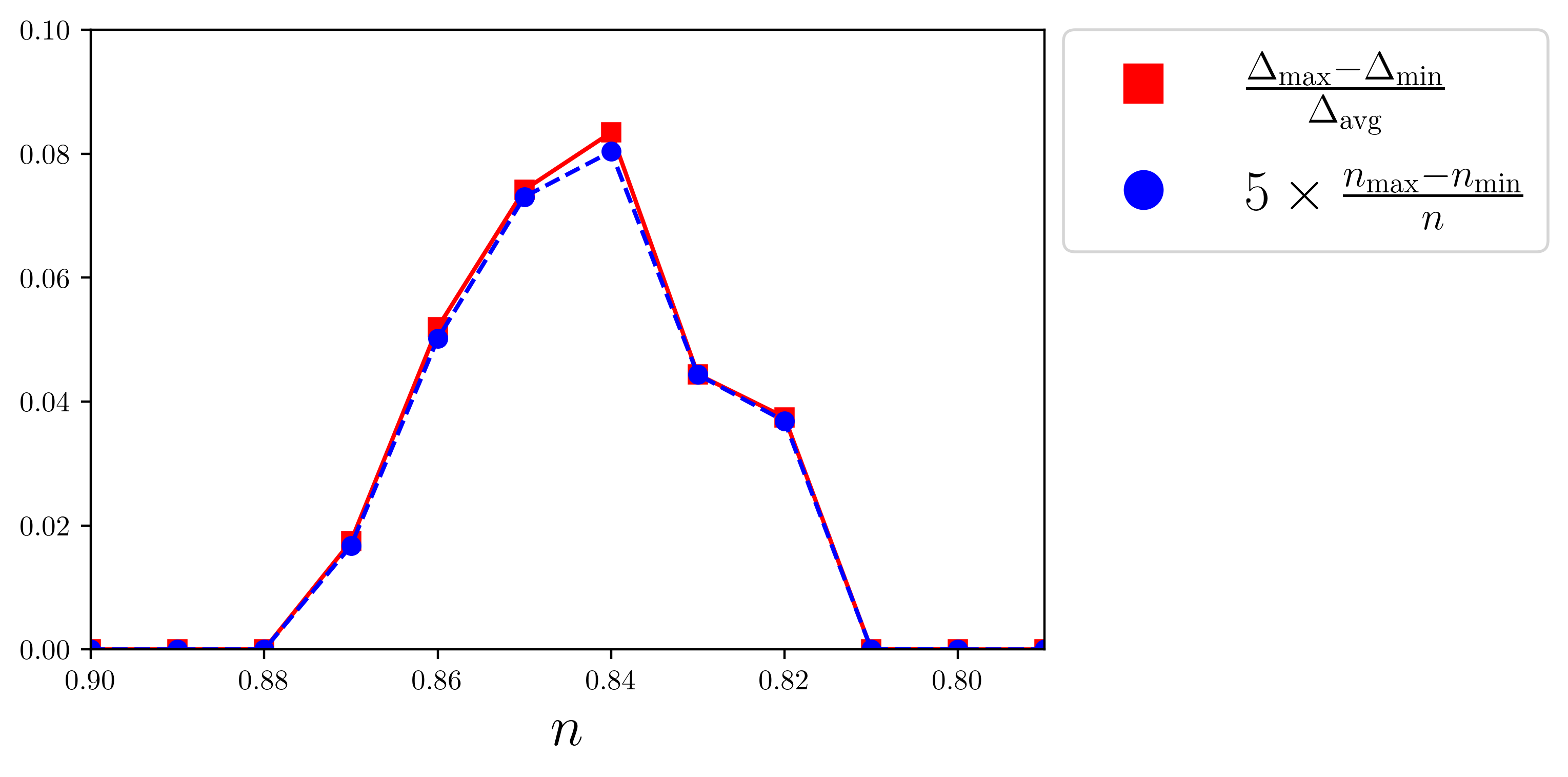}
 \caption{Relative modulations of the local density and the superconducting amplitude, calculated for different fillings at $T = 0.02t$.}
\label{fig: Modulations SC and Charge}
\end{figure}

In general, we find that the relative modulation of the charge order and the relative modulation of the superconducting order are proportional to each other, as illustrated in Fig.~\ref{fig: Modulations SC and Charge}.
In that figure, we present the relative modulations of filling and superconducting amplitudes along a density scan at a constant temperature $T = 0.02t$. We see that both modulations are zero in the spiral phase for $n \geq 0.88$. In the stripe phase, they are proportional to each other, with the superconducting modulation being roughly five times larger in this parameter region.
The superconducting amplitude typically has modulations of the order of a few percent, in agreement with recent numerical simulations \cite{Xu2024,Wietek2022,Baldelli2025}.
A similar modulated superconducting amplitude has been observed by STM experiments in cuprates \cite{Hamidian2016}.

In this regime, the spatial modulation of the superconducting gap is directly induced by the background density oscillations of the stripe order. It is important to distinguish these modulations from those characterizing Fulde-Ferrell-Larkin-Ovchinnikov (FFLO) states \cite{Matsuda2007}. While FFLO states are typically defined by a Cooper pair momentum that causes the order parameter to oscillate through zero, the modulations observed here represent a spatial ripple superimposed on a finite, non-zero mean value, a signature of superconductivity adapting to a pre-existing charge-inhomogeneous background. 


\subsection{40$\times$40 lattice} \label{Sec: Bigger Lattice}

To clarify which states remain stable in the thermodynamic limit, we have repeated our finite-size calculations on larger $40\times 40$ lattices. Due to the high computational cost, we focused these calculations on what we consider to be the most delicate region of the phase diagram: $n < 0.9$ and $T \leq 0.07t$.
\begin{figure}
\centering
 \includegraphics[width=0.49\textwidth]{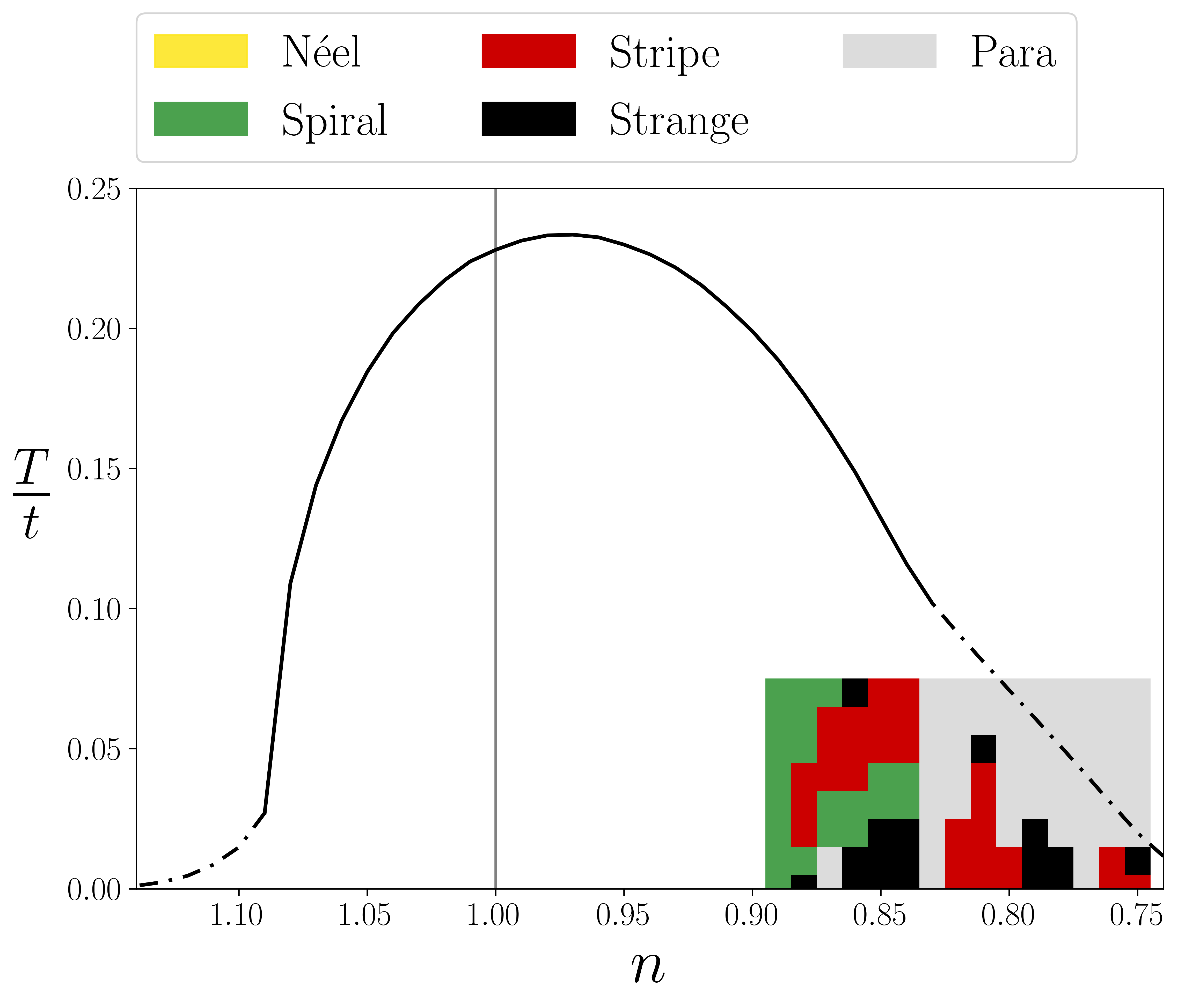}
 \caption{Magnetic phase diagram for the same parameters as in Fig.~\ref{fig: MagneticPhaseDiagram}, this time calculated on a $40 \times 40$ lattice. The critical temperature (black line) has been obtained in the thermodynamic limit as before. Due to the high numerical cost, ordered states have been computed only for $n < 0.9$ and $T \leq 0.07t$.}
\label{fig: PhaseDiagram4040}
\end{figure}
Our results for the magnetic phase diagram on the larger lattice are shown in Figure~\ref{fig: PhaseDiagram4040}. On the $40 \times 40$ lattice, the spiral phase remains stable at $n=0.89$. For higher doping values, we once again find stripe order, but this time it is adjacent to a region of spiral order at lower temperatures. The ``strange order phases'' seen in this calculation are likely due to convergence and finite-size issues on the large lattice \cite{Scholle2023}, and we expect them to vanish in the thermodynamic limit.

An important finding on this larger lattice is the complete suppression of magnetic order near van Hove filling at $n \approx 0.83$. At $T=0$, this feature has already been observed in prior work using the fRG + MF method, where magnetic order was restricted to spiral \cite{Yamase2016}. We confirmed this result by solving the gap equations restricted to magnetic spiral order and superconductivity in the thermodynamic limit at zero and finite temperature. This revealed a vanishing magnetization at $n=0.83$ for all temperatures, and thus a pure $d$-wave superconducting phase. For larger dopings, a dome of magnetic order emerges in these thermodynamic limit calculations, similar to the results from the unrestricted $40\times40$ lattice. This indicates that the reemerging magnetic region below van Hove filling obtained in the finite size calculations survives in the thermodynamic limit.

We note that in the calculations on a $28 \times 28$ lattice, we did find magnetically ordered states even at van Hove filling. However, as one can see in Fig.~\ref{fig: Magnetization}, the magnetization around $n = 0.83$ is rather weak. It seems that the finite lattice size stabilizes a weak magnetic order that disappears in the thermodynamic limit.


\subsection{Instability of the spiral order} \label{Sec: Susceptibilities}

In plain Hartree-Fock theory, the instability of the spiral state upon increasing the hole doping is signaled by a divergence of the charge susceptibility \cite{Scholle2024}. This divergence leads to a transition, first to a multi-spiral magnetic order and eventually to stripe order. To investigate this phenomenon in the presence of superconductivity, we have calculated the charge susceptibility $\chi(\bq)$ for coexisting spiral and superconducting orders, using the method described in Ref.~\cite{Vilardi2025SCstar}.

To this end we restrict the magnetic part of the renormalized mean-field Hamiltonian \eqref{eq: BdG_Hamiltonian} to spiral order, which allows us to perform calculations directly in the thermodynamic limit. The spiral order can be described by a simple $2 \times 2$ matrix Hamiltonian \cite{Kampf1996,Bonetti2022}.
This is extended to include superconductivity by using a Nambu spinor in momentum space
$\Psi_\mathbf{k} = ( c_{\bk\uparrow}, c_{\bk+\bQ\downarrow},
 c^\dagger_{-\bk-\bQ\uparrow}, c^\dagger_{-\bk\downarrow})$ \cite{Yamase2016}.
The Hamiltonian can then be written as
\begin{equation}
 H^{\text{MF}}_{\text{sp+sc}} =
 \sum_\bk \Psi_\mathbf{k}^\dagger \mathcal H_{\bk}\Psi_\mathbf{k} \, ,
\end{equation}
where
\begin{equation}
 \mathcal H_{\bk} =
 \left( \begin{matrix}
 \xi_{\bs{k}} & A & 0 & \Delta_\bs{k}  \\
 A & \xi_{\bs{k}+\bs{Q}} & -\Delta_{-\bs{k}-\bs{Q}} & 0 \\
 0 & -\Delta_{-\bs{k}-\bs{Q}} & - \xi_{-\bs{k}-\bs{Q}}  & -A \\
 \Delta_\bs{k} & 0 & -A & - \xi_{-\bs{k}}
 \end{matrix} \right) \, .
\end{equation}
where we defined $\xi_\bk = \epsilon_\bk - \mu$. 
Only for this section, we restrict our calculation to a $d$-wave pairing gap,
$\Delta_\bk = \Delta (\cos k_x - \cos k_y)$,
while the magnetic gap, $A$, is momentum independent.
The Green's function for this system is then given by
\begin{equation} \label{eq: GreensFunction Spiralstate}
 \mathcal{G}^{-1}_\bk(i\nu_n) = i\nu_n \mathbb{1}_{4\times4} - \mathcal H_\bk \, .
\end{equation}

Following the procedure detailed in Ref.~\cite{Vilardi2025SCstar}, we define vertex matrices to construct the relevant fermion bilinears in the coexisting phase.
\begin{subequations}
\begin{align}
 M^{a} &= \left(\begin{matrix}
 \sigma^a & 0 \\
 0 & - \left(\sigma^a\right)^{\mathrm{T}} \end{matrix}\right), \\
 h &= \left(\begin{matrix}
 0 & i\sigma^2 \\
 -i\sigma^2 & 0 \end{matrix}\right), \\
 \Phi &= \left(\begin{matrix}
 0 & \sigma^2 \\
 \sigma^2 & 0 \end{matrix}\right) \, ,
\end{align}
\end{subequations}
with $a=0,1,2,3$, where $\sigma^0 = {\rm diag}(1,1)$, and $\sigma^1,\sigma^2,\sigma^3$ are the Pauli matrices.
These matrix vertices allow us to define a generalized susceptibility
\begin{equation}
 \chi^{ab}(q) = \int_{k,k'}
 \langle \Psi^\dagger_{k+q} \Gamma^a \Psi_k \Psi^\dagger_{k'-q} \Gamma^b \Psi_{k'} \rangle \, .
\end{equation}
Here $k$, $k'$, and $q$ are combined momentum and frequency variables, and $\int_k$ is a shorthand notation for the Matsubara frequency sum and the momentum integral, with the appropriate prefactors $T$ and $(2\pi)^{-2}$ included. The components of the vertex
$\Gamma = (\Gamma^0,\dots,\Gamma^5) = (M^0,M^1,M^2,M^3,h,\Phi)$ capture all possible charge, spin, and pairing correlations.
In particular, $\chi^{00}$ is the charge susceptibility, and $\chi^{ab}$ with $a,b \in \{1,2,3\}$ are the various components of the magnetic susceptibility.

In the RPA approximation, the susceptibility is given by
\begin{equation} \label{eq: rpa}
 \chi(\bq,\omega) = \frac{\chi_0(\bq,\omega)}{\mathbb{1} - 2{\cal U} \chi_0(\bq,\omega)} \, ,
\end{equation}
where $\chi(\bq,\omega)$, $\chi_0(\bq,\omega)$, and ${\cal U}$ are $6\times6$ matrices, with
${\cal U} = \mathrm{diag}(-U^m,U^m,U^m,U^m,U^p,U^p)$.
The matrix elements of the bare susceptibility $\chi_0(\bq,\omega)$ can be obtained from the Green's function in the form
\begin{eqnarray}
 \chi^{ab}_0(\bs{q},\omega) &=&
 -\frac{1}{8}\int_{\bs{k}} f^a_\bs{k}f^b_\bs{k} \, T\sum_{\nu_n} \Tr
 \big[ \Gamma^a \mathcal{G}_{\bk+\bq}(i\nu_n+i\Omega_m) \nonumber \\
 && \times \Gamma^b \mathcal{G}_\bk(i\nu_n) \big]
 \big|_{i\Omega_m \rightarrow \omega +i0^+} \, ,
\end{eqnarray}
with the form factors $(f^0_\bk,\dots,f^5_\bk) = (1,1,1,1,d_\bk,d_\bk)$.

\begin{figure}
\centering
 \includegraphics[width=0.49\textwidth]{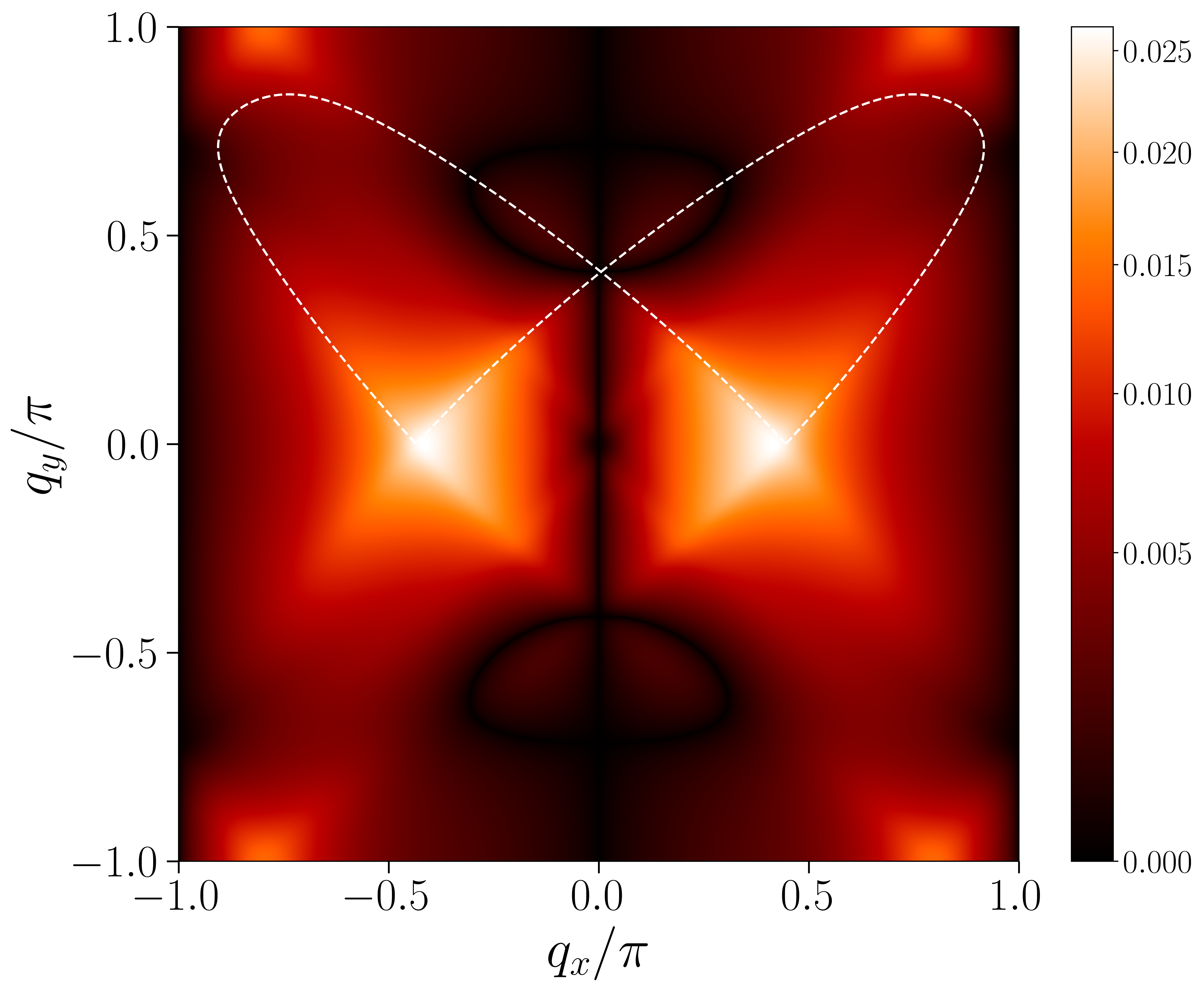}
 \caption{Momentum dependence of $|\chi_0^{12}(\bq)|$ for $n = 0.86$ and $T = 0.04t$ in units of $\frac{1}{t}$. The pronounced maxima on the $q_x$ axis are similar to those driving the instability of the pure spiral state toward charge order \cite{Scholle2024}.
 In the upper half of the Brillouin zone, we retrace the $2k_F$-lines with dashed white lines.}
\label{fig: Bubble}
\end{figure}

In Ref.~\cite{Scholle2024} it was shown that the momentum dependence of the off-diagonal component $\chi_0^{12}(\bq)$ plays a crucial role in driving the instability of the spiral state towards a state with spin and charge order. This structure is linked to the shape of the Fermi pockets in the pure spiral state. We find a similar behavior even in the presence of superconductivity. As shown in Figure \ref{fig: Bubble}, for $n = 0.86$ the absolute value of $|\chi_0^{12}(\bq)|$ displays pronounced peaks on the $q_x$ axis.
These peaks are situated at crossing points of ``$2k_F$-lines'', which are formed by the nesting vectors connecting points with parallel tangents on the Fermi surface of the pockets \cite{Holder2012}.
Even though the pairing gap truncates the Fermi surface of the spiral pockets, leaving only nodal points in the Brillouin zone, the momentum structure of $\chi_0^{12}(\bq)$ does not change very much, compared to the normal (non-superconducting) state. In Fig.~\ref{fig: Bubble}, we retrace the $2k_F$-lines of the spiral state we find for $n=0.86, T=0.04t$, neglecting its superconducting order parameter, in the upper half of the Brillouin zone. One can see that the maximum of $|\chi_0^{12}(\bq)|$ occurs at the crossing point of the two lines.

To further analyze the transition from spiral to stripe order, we directly study the divergence of the charge susceptibility component, $\chi^{00}(\bq)$.
\begin{figure}
\centering
 \includegraphics[width=0.49\textwidth]{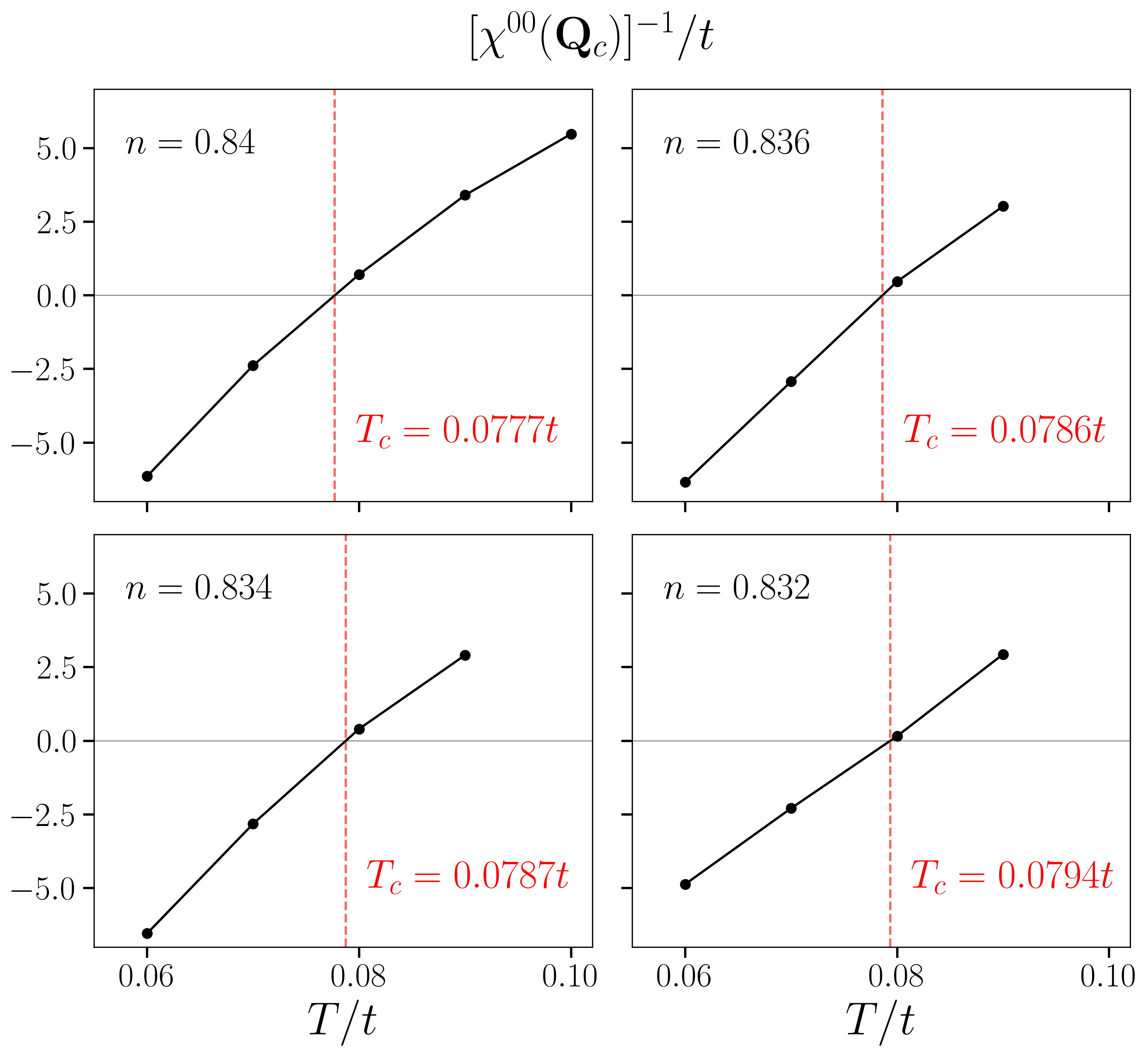}
 \caption{The minimum of $[\chi^{00}(\bq)]^{-1}$ for four different fillings, as a function of temperature. We marked with the red dashed line the temperature at which the susceptibility diverges. For higher temperatures the spirals are stable. We see that $T_c$ saturates around $0.08t$ as we approach van-Hove filling at $n = 0.830$ from above.}
\label{fig: ChargeSusceptibilities}
\end{figure}
In Figure \ref{fig: ChargeSusceptibilities} we plot the inverse charge susceptibility at its minimum, $[\chi^{00}(\mathbf{Q}_c)]^{-1}$, as a function of temperature for various fillings slightly above the van Hove filling. Here, we defined $\mathbf{Q}_c$ as the momentum where the inverse charge susceptibility has its minimum. A crossing of zero indicates a divergence in the susceptibility. At the van Hove filling itself, $n=0.83$, the spiral order is not observed and is replaced by a purely superconducting state.
We find that the critical temperature for the spiral instability remains nearly constant upon approaching the van Hove filling from above, saturating at approximately $T=0.079t$, before dropping abruptly to zero at the van Hove filling.
This temperature is notably lower than the critical temperature $T^*$ for the onset of magnetic order.

\begin{figure}
\centering
 \includegraphics[width=0.49\textwidth]{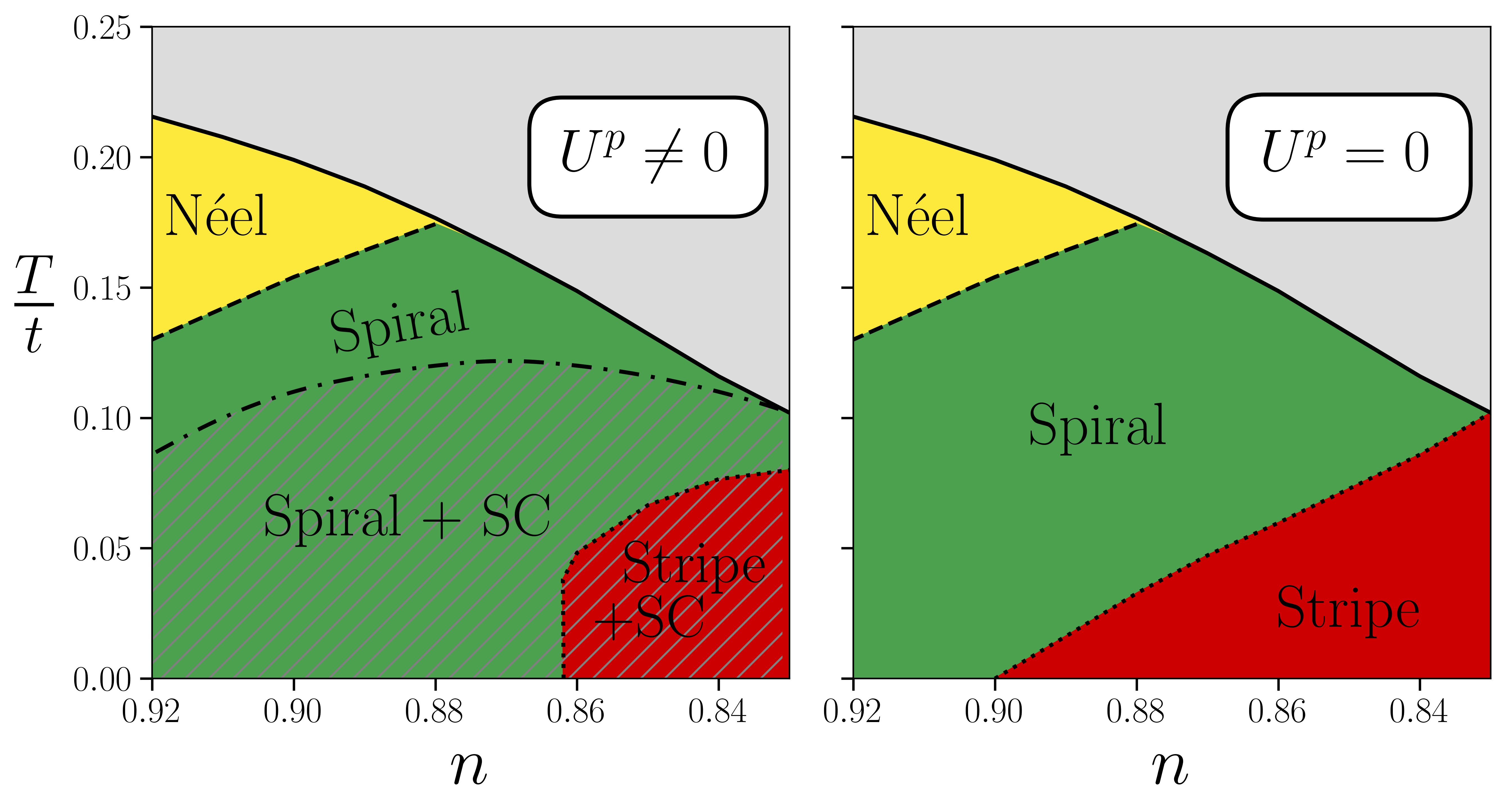}
 \caption{Comparison of the phase boundaries between the spiral and the stripe state.
 Left: with the effective interactions $U^m$ and $U^p$ as obtained from the fRG calculations. The dash-dotted line indicates the onset of superconductivity. Right: with $U^m$ as obtained from the fRG calculations, but with $U^p$ set to zero. Here, the stripe regime extends over a larger area of the phase diagram.}
\label{fig: Comparison Stripe Border}
\end{figure}
Figure \ref{fig: Comparison Stripe Border} illustrates the critical instability line that separates the spiral and stripe phases. The left panel shows this line for our full calculation with the pairing interaction, while the right panel presents the result for the artificial case with the pairing interaction set to zero ($U^p=0$). A comparison reveals that the spiral phase remains stable over a significantly wider doping range when the pairing interaction is included. In particular, at $T=0$, the spiral-to-stripe transition shifts from around $n=0.9$ in the absence of superconductivity to $n \approx 0.86$ in its presence. This indicates that the presence of superconductivity suppresses stripe order and instead favors a coexistence with spiral order.

For some fillings (for example, $n=0.84, 0.85, 0.86$), the inverse charge susceptibility in the spiral state initially decreases and crosses zero as temperature is lowered, indicating an instability, but then rises back above zero at even lower temperatures ($T < 0.03t$ for $n=0.86$, see App.~\ref{App: reentrance}).
This indicates that spiral order may be stable again in this low temperature region, or at least metastable. This hypothesis is supported by our $40 \times 40$ lattice calculations, where a small spiral phase reemerges at low temperatures inside the stripe-ordered dome (see Figure \ref{fig: PhaseDiagram4040}). Given the significant finite-size effects on the small energy difference between spiral and stripe states, determining which is the true ground state in this region would require calculations on even larger lattices.
We therefore refrain from computing and plotting the additional putative transition line associated with a possible low temperature reentrant spiral state.


\subsection{Special point: $t' = 0 , \, n = \frac{7}{8}$ and $T = 0$}

A special point in the large parameter space of the two-dimensional Hubbard model has become particularly popular for exact numerical simulations on finite lattices: the ground state ($T = 0$) for pure nearest-neighbor hopping ($t'=0$) at a density of $n = \frac{7}{8}$ \cite{Zheng2017, Jiang2020, Qin2020}. At strong coupling, strong evidence for an insulating stripe state has been found \cite{Zheng2017,Qin2020}.
We have analyzed this parameter set with the fRG+MF method, albeit with the same moderate interaction strength $U = 4t$ as we used for the previously presented results.
The corresponding effective interactions resulting from the fRG are $U^m = 2.159t$ and $U^p = 0.864t$.

We performed fRG+MF calculations for several different lattice sizes, namely, on $28\times 28$, $32\times 32$, and $40\times 40$ lattices. For the $28\times 28$ and $32\times 32$ lattices, we find a uniform superconducting solution without any charge modulations or magnetic order. However, for the $40\times 40$ lattice we find stripes with a sizable magnetization of $m = 0.06$ and an incommensurability $\eta = \frac{1}{20}$. These stripes coexist with (modulated) $d$-wave superconductivity of strength $\Delta = 0.061t$.
The free energy of this state is $\frac{F}{\mathcal{N}} = -1.19671t$, which is slightly lower than that of the optimal purely superconducting state on this lattice, $\frac{F^{\text{only SC}}}{\mathcal{N}} = -1.19648t$, obtained by enforcing $m = 0$ in each iteration. This suggests that the state of coexisting magnetism and superconductivity is likely stable in the thermodynamic limit, however it is energetically very close to the purely superconducting case. Therefore finite size effects can easily suppress the magnetization at this point.


\subsection{Thermodynamic Limit}\label{sec: Thermodyn Limit}

\begin{figure}
\centering
 \includegraphics[width=0.49\textwidth]{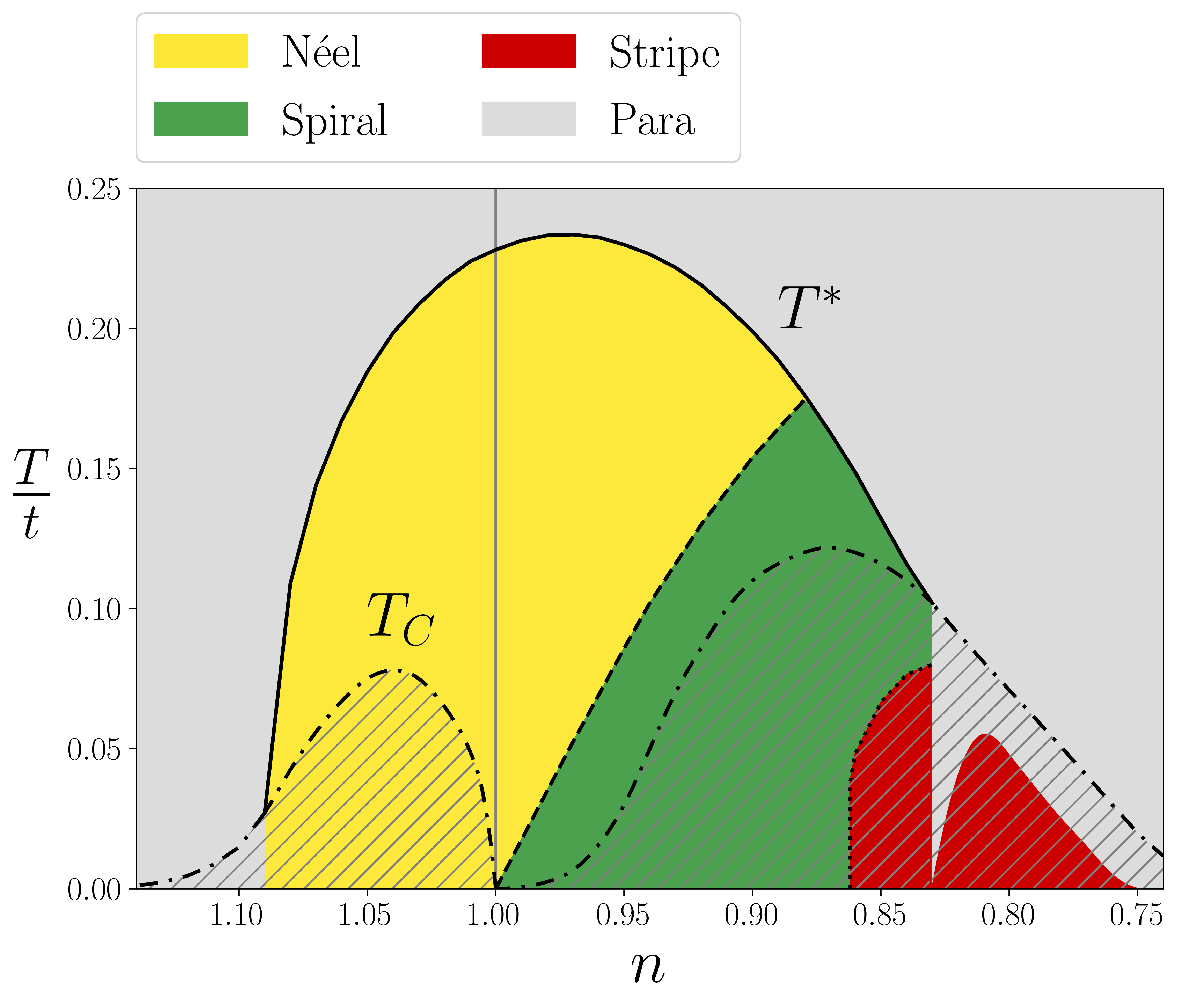}
 \caption{$(n,T)$ phase diagram for $U = 4t$ and $t^\prime =-0.2t$ in the thermodynamic limit. All the black lines have been computed directly in the thermodynamic limit. The solid line indicates the onset of magnetic order. The dash-dotted lines show the onset of superconductivity, and the shaded areas the superconducting regions. The dotted line indicates the instability of the spiral states, as signaled by a diverging charge susceptibility. Near the spiral region, the region labeled as ``stripe'' may also contain a non-collinear charge ordered state such as the multi-spiral state in Ref.~\cite{Scholle2024}. The boundary of the small stripe-ordered dome at low densities could only be estimated from the finite size calculations.}
 \label{fig: PhaseDiagramTDL}
\end{figure}
In Fig.~\ref{fig: PhaseDiagramTDL} we present our expectation for the phase diagram of the two-dimensional Hubbard model with $U = 4t$ and $t' = -0.2t$ in the thermodynamic limit. It is based on the insights gathered from our unrestricted real-space calculations and the restricted calculations performed directly in the thermodynamic limit. A similar extrapolation has been performed previously for a plain Hartree-Fock calculation \cite{Scholle2023}, where superconductivity is not captured.

The critical temperatures for the onset of spontaneous symmetry breaking have been obtained directly from the fRG flow (in the thermodynamic limit). The solid black line indicates the critical temperature where the divergence of the two-particle vertex occurs in the magnetic channel, whereas a divergence in the pairing channel is indicated by the dash-dotted black line. In the hole-doped regime, we have calculated the transition between N\'eel and spiral order directly in the thermodynamic limit, as in Ref.~\cite{Bonetti2022gauge}, and obtained good agreement with the finite size results in Fig.~\ref{fig: MagneticPhaseDiagram}. Also the critical temperatures of the superconducting state have been computed with both approaches in very good agreement, both on the electron- and hole-doped sides. The critical temperature rises more steeply on the electron-doped side, as already seen in Ref.~\cite{Metzner2019}, but we find ultimately higher critical temperatures and larger values of the pairing gap on the hole-doped side of the phase diagram, in agreement with the same trend in the cuprates.
In the stripe phase, the pairing gap is modulated with the same wave vector as the charge order, as analyzed in Fig.~\ref{fig: Modulations SC and Charge}. This modulation is typically of the order of a few percent.
A spatially modulated pairing gap has also been observed experimentally by scanning tunneling microscopy in some cuprates \cite{Hamidian2016}.

Fluctuations of the superconducting order parameter would actually degrade the superconducting transition temperature obtained in our mean-field theory to a crossover scale for pairing, while the true transition temperature is lower and associated with a Kosterlitz-Thouless transition \cite{Metzner2019, Vilardi2020}.
The reduction of the Kosterlitz-Thouless temperature with respect to the pairing temperature is particularly pronounced for low electron doping \cite{Metzner2019}.

The dotted line in Fig.~\ref{fig: PhaseDiagramTDL} indicates where the charge susceptibility of the spiral state diverges in the thermodynamic limit, signaling the transition towards stripe order. This transition roughly coincides with the onset of stripe order in our real-space calculations, shown in Figs.~\ref{fig: MagneticPhaseDiagram} and \ref{fig: PhaseDiagram4040}.
The divergence of the susceptibility indicates a second order transition of the spiral state to an intermediate phase which will eventually turn into collinear stripe order, as in the pure Hartree-Fock calculation in the absence of superconductivity \cite{Scholle2024}. In Appendix~\ref{App: reentrance} we show that spiral order may reemerge for low temperatures below the dotted line.

At van Hove filling ($n=0.83$), the momentum space calculation yields a uniform superconducting state without any spin or charge order, in agreement with (unrestricted) real-space calculations on a $40\times40$ lattice, but in contrast to the coexisting stripe order found on the smaller $28\times28$ lattice.
In the regime below van Hove filling ($n<0.83$) distinct solutions found with the two methods have extremely close energy values, making the extrapolation to the thermodynamic limit difficult. For instance, for $n=0.80$ and $T = 0$, the momentum space calculation in the thermodynamic limit yields a purely superconducting solution which is only $10^{-7}t$ higher in energy than the solution with coexisting circular spiral order. On the $40 \times 40$ lattice, where for this parameter set we find stripe order coexisting with superconductivity, the energy gain is $6\cdot10^{-6}t$, which is larger than for the spiral order, yet still very close to the purely superconducting solution. Hence, in this regime the majority of the energy gain is due to the superconducting order parameter, rather than magnetism, as observed already in Ref.~\cite{Yamase2016}.
The spiral states found for densities $0.80 < n < 0.83$ within the momentum space calculations exhibit a diverging charge susceptibility, confirming that stripe order is the preferred magnetic pattern in this regime, in agreement with the results of the real-space calculations on $40\times40$ lattice (see Fig.~\ref{fig: PhaseDiagram4040}).
We schematically sketched this fragile second magnetic dome below van Hove filling, retracing the envelope of the magnetic regions found on the $40\times 40$ lattice.


\section{Conclusions} \label{Sec: Conclusion}

We have performed a renormalized mean-field study of the two-dimensional Hubbard model to explore the interplay and possible coexistence of magnetic, charge, and superconducting order. Our approach improves on conventional mean-field theory by using a functional renormalization group flow to replace bare interactions in the mean-field equations by renormalized effective interactions. The strength of magnetic interactions is thereby reduced, while pairing interactions with $d$-wave symmetry are being generated.
We have constructed completely unrestricted solutions of the renormalized mean-field Hamiltonian in real space on various large finite lattices ($28 \times 28$, $32 \times 32$, and $40 \times 40$), and complemented these by momentum space solutions in the thermodynamic limit, where the magnetic order was restricted to spiral or N\'eel order.

Explicit results have been presented for an intermediate bare Hubbard interaction $U=4t$, and a sizable next-to-nearest neighbor hopping amplitude $t'=-0.2t$ has been chosen in most of the calculations, to mimic the band structure of cuprates.
The $(n,T)$ phase diagram in Fig.~\ref{fig: PhaseDiagramTDL} summarizes our results for this parameter set. There are essentially three types of magnetic order: N\'eel, circular spiral, and stripe. Except for half-filling, the magnetic order always coexists with superconductivity at sufficiently low temperatures. The pronounced electron-hole asymmetry is due to the presence of $t'$.

At half-filling and on the electron-doped side ($n>1$), the magnetic order is exclusively of N\'eel type, with coexisting $d$-wave superconductivity at sufficiently low temperatures for any $n \neq 1$.

On the hole-doped side ($n<1$), the N\'eel state first transitions into a spiral state upon increasing the doping, followed by stripe order. N\'eel order persists in the hole-doped regime only at finite temperatures. At sufficiently low temperatures, the magnetic order is accompanied by superconductivity in the entire hole-doped regime. While the pairing gap is spatially uniform in the N\'eel and spiral regimes, it becomes spatially modulated in the stripe region, following the pattern of the charge order. The strength of the modulation is a few percent, in agreement with experimental observations by scanning tunneling microscopy in some cuprates \cite{Hamidian2016}.
Comparing the complete phase diagram to a phase diagram where superconductivity is artificially switched off (right panel of Fig.~\ref{fig: Comparison Stripe Border}), one can see that pairing reduces the region with stripe order (and hence charge order) in favor of spiral magnetic order. Experimentally superconductivity can be suppressed by a strong magnetic field, which can indeed promote the emergence of charge order in some cuprates \cite{Wu2011,Chang2012}.

At van Hove filling the magnetic order is completely suppressed by superconductivity, even in the ground state, so that a purely superconducting phase remains. This peculiarity has already been found previously in a renormalized mean-field calculation restricted to N\'eel and spiral order \cite{Yamase2016}. In the unrestricted real-space calculation, this feature became apparent only on our largest system, the $40\times40$ lattice.

For hole dopings beyond the van Hove filling, we find a region primarily dominated by superconductivity, with some areas of coexisting stripe order. By comparing the energies of all solutions, it turns out that the superconducting state provides the most significant energy gain in this regime, while the stripe order on top of it lowers the energy only by a tiny amount, suggesting that this coexisting spin-charge order is very fragile.
Restricting the magnetic order to spiral order, a similarly fragile coexistence has been found previously, and referred to as ``gossamer magnetism'' \cite{Yamase2016}.

We have also analyzed the ground state order at the special density $n = \frac{7}{8}$ for the Hubbard model with pure nearest-neighbor hopping ($t'=0$), a special case which has been particularly popular for exact numerical simulations on finite lattices.
While these simulations yield strong evidence for an insulating stripe state at strong coupling \cite{Zheng2017,Qin2020}, our calculations for the moderate interaction strength $U = 4t$ yield a superconducting stripe state. It is plausible that at stronger coupling the stripe order opens a large charge gap which prevents pairing.

The perturbative renormalization group flow used to compute the effective interactions limits our technique to weak or moderate interactions. A leap to strong coupling is possible, but numerically quite demanding, by starting the renormalization group flow from a dynamical mean-field solution, so that the strong local correlations are taken into account non-perturbatively from the very beginning \cite{Taranto2014, Vilardi2019}.

Order parameter fluctuations have been completely neglected in our work. It is clear that they restore the SU(2) spin rotation symmetry and the U(1) charge symmetry at any finite temperature. One could take them into account a posteriori in the framework of gauge theories of fluctuating order, where the electron is fractionalized into a chargon, which may break pseudospin and pseudocharge symmetries, and a fluctuation part which restores the physical symmetry \cite{Scheurer2018, Sachdev2019, Sachdev2019review, Bonetti2022gauge}.


\section*{Acknowledgments}
We are very grateful to A.~Chubukov, P.~Forni, A.~Georges, H.~M\"uller-Groeling, S.~Sachdev, and Shiwei Zhang for valuable discussions.
P.M.B.\ acknowledges support by the German National Academy of Sciences Leopoldina through Grant No.\ LPDS 2023-06, by the Gordon and Betty Moore
Foundation’s EPiQS Initiative Grant GBMF8683, and by the NSF Grant DMR-2245246.


\begin{appendix}

\section{Numerical details of mean field loop} \label{App: Numerical Details}

We have to define a chemical potential $\mu$ to enter in Eq.~\eqref{eq: Hm}. For that, we start with an initial chemical potential and adjust it in each iteration via
\begin{equation}
    \mu_{\text{new}} = \mu_{\text{old}} - \alpha ({\bar n} - n)\,,
\end{equation}
with the current average density ${\bar n} = {\cal N}^{-1} \sum_j \langle n_j \rangle$ and $\alpha > 0$ so that the chemical potential increases (decreases), when the current filling is smaller (bigger) than the desired filling. In standard Hartree-Fock calculations with a Hamiltonian that is not of BdG type, one can directly fix the chemical potential in each iteration self-consistently \cite{Scholle2023}. Here, this is not possible, since Hamiltonian~\eqref{eq: BdG_Hamiltonian} does not depend trivially on $\mu$, which is why we chose this iterative approach. The chemical potential and the filling converge to the final correct values.
In our calculations, we have mainly set $\alpha = 0.35$. 
We furthermore employ linear mixing of roughly 50\% of the expectation values and the chemical potential in each iteration.

We consider a state as converged, if all the parameters $A_{j\alpha}$ and $\Delta_{j\alpha}$ change by less than $10^{-8}t$ between two iterations, or after 3000 iterations.
Further lowering the threshold below $10^{-8}t$ does not alter the results.
The second convergence criterion was introduced to guarantee eventual termination of the computation. In practice, 3000 iterations are typically sufficient to reach a well-defined state that reflects the same type of magnetic order as the fully converged solution. Furthermore, by performing multiple runs for each $(n,T)$ point with different initial conditions, we are confident that the identified magnetic order corresponds to the true global minimum for each parameter set.  


\section{Possible reentrance of the spiral state} \label{App: reentrance}

\begin{figure}
\centering
 \includegraphics[width=0.49\textwidth]{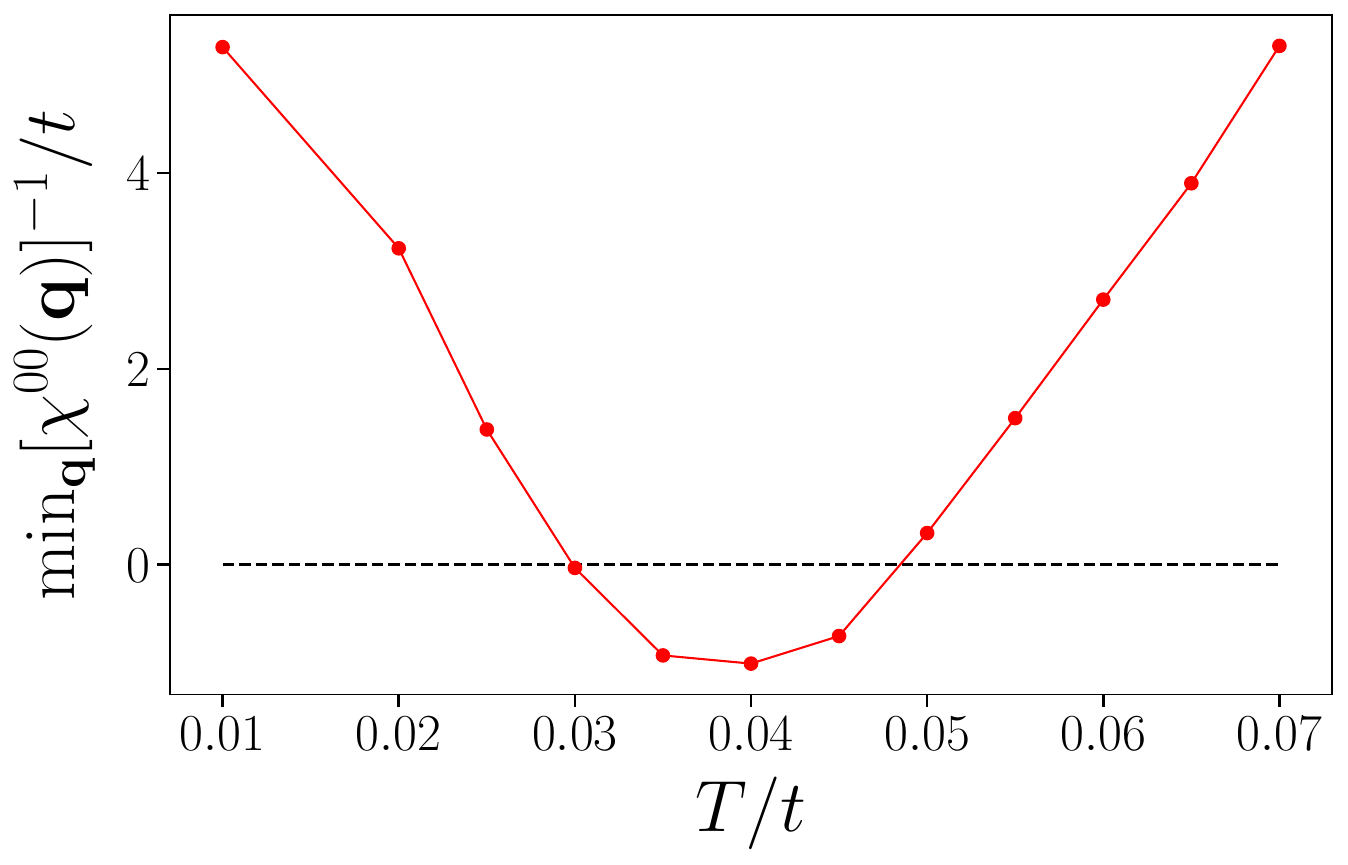}
 \caption{Minimum in $\bq$ of the inverse charge susceptibility as a function of $T$ for $n = 0.86$.}
\label{fig: ChargeSusceptibilities_n086}
\end{figure}
As described in Sec.~\ref{Sec: Susceptibilities}, we determined the border of the spiral regime by a divergence in the charge susceptibility in the spiral state. In Fig.~\ref{fig: ChargeSusceptibilities_n086}, we show the minimum in $\bq$ of the inverse charge susceptibility as a function of $T$ for $n = 0.86$. As we lower the temperature, we first see a positive charge susceptibility $\chi^{00} > 0$ for $T > 0.05t$, followed by a regime where $\chi^{00}(\bq) < 0$ for some momenta in the Brillouin zone, indicating that the spiral order is unstable. As we keep lowering the temperature below $T = 0.03t$, the charge susceptibility becomes positive again, indicating that the spiral order is at least meta stable in this regime. To determine whether spirals or stripes yield the lowest free energy in this low temperature regime in the thermodynamic limit, one would need to compare their energies on very large lattices, since the finite-size effects on the lattices we consider here are of the order of the energy difference between spirals and stripes for this parameter set.

\end{appendix}


\bibliography{main.bib}

\end{document}